\documentclass[12pt,sort&compress]{article}
\pdfoutput=1

\usepackage[dvips]{color}
\usepackage{amsmath}
\usepackage{graphicx} 
\usepackage[dvips]{color}
\usepackage{graphicx}
\usepackage{tabularx}
\usepackage{subfig}
\usepackage{cite}
\usepackage[section]{placeins}
\usepackage{hyperref}

\newcommand{\signum}{\mathrm{sign}}
\newcommand{\kernel}{\mathcal{K}}
\newcommand{\Ai}{\mathrm{Ai}}
\newcommand{\pfaffian}{{\mathrm{Pf}}}
\newcommand{\cbar}{\bar{c}}
\newcommand{\proba}{\mathrm{Prob}}

\begin{document} 
\title{Log-gamma directed polymer with one free end via  coordinate Bethe Ansatz}
\author{Pascal Grange\\
Department of Mathematical Sciences\\
 Xi'an Jiaotong-Liverpool University\\
111 Ren'ai Rd, 215123 Suzhou, China\\
\normalsize{{\ttfamily{pascal.grange@xjtlu.edu.cn}}}}

\date{}
\maketitle
\vspace{1cm}
\begin{abstract}
 The discrete polymer model with random Boltzmann weights  with homogeneous inverse 
 gamma distribution, introduced by Sepp\"al\"ainen, is studied in the case of a polymer 
 with one fixed and one free end. The model with two fixed ends has been integrated by Thiery
 and Le Doussal, using coordinate Bethe Ansatz techniques and an analytic-continuation 
 prescription. The probability distribution of the free energy has been obtained through the replica method,
  even though the  moments of the partition sum do not exist at all orders due to the fat tail in the
 distribution of Boltzmann weights. To extend this approach to the polymer with one free end, we
 argue that the contribution to the partition sums in the thermodynamic limit is localised on parity-invariant
   string states. This situation is analogous to the 
 case of the continuum polymer with one free end, related to the Kardar--Parisi--Zhang equation
 with flat boundary conditions and solved by Le Doussal and Calabrese. The expansion of the 
 generating function of the partition sum in terms of numbers of strings can also be transposed
 to the log-gamma polymer model, with the induced Fredholm determinant structure.
 We derive the large-time limit of the rescaled cumulative distribution 
 function, and relate it to the GOE Tracy--Widom distribution. The derivation is conjectural in the sense 
 that it assumes completeness of a family of string states,  and expressions of their norms, already 
  useful in the fixed-end problem, and extends heuristically  the order of moments of the partition sum
 to the complex plane. 
\end{abstract}

\newpage

\tableofcontents




\section{Introduction and conclusions}

The application of  the methods  of integrable 
systems to the continuum Kardar--Parisi--Zhang (KPZ) equation \cite{KPZ,KPZRep}, a model 
 of one-dimensional growth of an interface  in the presence  of noise, 
  led to an exact solution. In particular, the height field of the interface was characterised by its probability 
 density, by means of the Cole--Hopf mapping which relates the height 
 field to the free energy of a directed-polymer model. The time evolution of the 
 integer moments of the partition sum of this model, in the replica approach, was mapped to the 
 one-dimensional Lieb--Liniger model of interacting bosons \cite{LL,BrunetDerrida1,BrunetDerrida2}, which is solvable by Bethe Ansatz methods 
 \cite{GaudinTraduction}. The collection of exact expressions for these moments yielded the probability density 
 of the rescaled free energy for the most studied classes of boundary conditions  
 for the KPZ equation. The  moments were rewritten in terms of the Airy function, as they grow
 proportionally the exponential of the cube 
  of their order.  The resulting kernels allowed to relate the
   probability density of the free energy at large times to the Tracy--Widom \cite{TW,GOEKernel}
 distributions of the largest eigenvalue of large Gaussian random matrices. Moreover, the generating function 
 of moments was expressed as a Fredholm determinant for several classes of boundary conditions.\\

 While this research program has been completed 
 for continuum models, yielding insights on the KPZ universality class \cite{Rosso,Dotsenko,DotsenkoKlumow,CalabreseLeDoussalFlat,CalabreseLeDoussalLong,IS1,IS2,
GueudreLeDoussal,131,132,133,Spohn1,Spohn2,Spohn3,Spohn4,Spohn5,SasamotoSpohn1,Quastel,Corwin11,LeDoussal,CalabreseLeDoussalQuench,sineGordon},
 an alternative approach is the study of discrete models of directed polymers (or integrable particle systems or growth models),
 whose solutions can yield results for the continuum model by taking a suitable limit (see \cite{theseThiery} for a recent review,
 and \cite{Povolotsky,TLDSquare,ThieryStat,TLDBeta,Yor,HigherSpin,strictWeak,betaRandom} for more families of models and their classification,
 and \cite{ISDuality,BCS14} for rigorous solutions involving the replica approach).
  In particular, the log-gamma polymer model introduced by Sepp\"al\"ainen \cite{logGammaDef},
 defined by the distribution of Boltzmann weights on a lattice (with a fat-tailed  
  distribution of weights characterised by an exponent denoted by $\gamma$), 
 was solved by combinatorial methods \cite{Corwin}, and the generating function was 
 related to a Fredholm determinant \cite{Borodin}.\\

 On the other hand, a crucial step towards the extension  of the Bethe Ansatz replica approach to discrete 
 models was taken by Thiery and Le Doussal in \cite{logGammaTLD}, where the fixed-end 
 log-gamma model was solved using the coordinate Bethe Ansatz approach due to Brunet, yielding eigenfunctions of the 
 replica transfer matrix. In particular, the tangents of the rapidities, which are 
  used to express these eigenfunctions,  are regularly spaced on segments of the complex plane
 in the thermodynamic limit. This reproduces the so-called string structure that rapidities
  exhibit in the Bethe Ansatz solution of the Lieb--Liniger model.\\

 However, the fat tail of the distribution of Boltzmann weights is an obstacle to the 
  replica approach to the log-gamma polymer, which makes use the moments of the partition sum at all orders,
 whereas these moments diverge at orders larger than the parameter $\gamma$. An analytic-continuation
 prescription inspired by the Mellin representation of the exponential function
 was given in \cite{logGammaTLD} to express the generating function, given a formal expression 
 of integer moments of the partition sum. This led to an expression of the generating function as a Fredholm determinant.
 At large times  the rescaled cumulative distribution function of the free energy of the log-gamma 
directed polymer with fixed ends, reduces to the Gaussian Unitary Ensemble (GUE) Tracy--Widom distribution. This completed the 
 replica Bethe Ansatz program for the log-gamma polymer with fixed ends, whose continuum limit 
 can be related to the KPZ problem with droplet boundary conditions. This paper aims to 
 extend this approach to the large-time limit of  the log-gamma polymer with one free end and 
one fixed end (whose continuum limit corresponds to the KPZ problem with flat boundary conditions). The results are conjectural 
 in the sense that the same assumptions are made as in \cite{logGammaTLD}, including
 the completeness of the string solutions in the thermodynamic limit, the expression of the norms 
 of these states, and the  analytic continuation to complex values of the number of replicas. Moreover, the symmetry 
 argument used to restrict the calculation to parity-invariant states is only supported by checks in
 special cases. \\

The outline of the paper is as follows. In Section 2 the log-gamma polymer 
 model is reviewed. Notations are introduced for the rescaled free energy,
 and the analytic-continuation prescription  is motivated by the 
 divergence of the moments at time zero, which occurs  in the free-end case
 just as in the fixed-end case. The partition sum of the free-end model takes the form of
  a sum over all the fixed-end sectors.  Expressing the moments  in terms
 of the  orthogonal Bethe  eigenfunctions (assumed to form a complete set of states) yields a sum
 of quantities known from the fixed-end solution of  \cite{logGammaTLD}, weighted by overlaps between eigenfunctions
 and a uniform wave function on the lattice. These overlaps  are therefore the only quantities left 
 to compute. In Section 3 we express these overlaps after arguing that
 the phase-space integration localises on parity-invariant systems of string states. 
 String states appeared crucially for the continuum 
 polymer model with one free end in the derivation of Calabrese and Le Doussal in 
 \cite{CalabreseLeDoussalFlat,CalabreseLeDoussalLong},
 and were reobtained and interpreted by means of  a symmetry argument in \cite{CalabreseLeDoussalQuench}
 based on the results of \cite{deNardis} involving {\emph{half-strings}}. The 
 fact that the string states in the  thermodynamic limit of the log-gamma  model are obtained
 from the rapidities by an odd mapping relates the two situations. In Section 4 we work 
 out the first  term in the string expansion of the generating function, which induces the
  centering and  scaling of  the free energy in terms of the parameter $\gamma$. After analytic continuation, the large-time limit  
of this term is  studied using a saddle-point method and the Airy function, and found to be independent of the 
 parameter $\gamma$. Numerical checks are reported in the final section, based on direct 
 simulations of the system with $10^4$ samples, at time $t=4,096$, and for $\gamma=3$ (a value for which the 
 divergence of moments is especially severe).  
 It should be noted that an alternative (rigorous, non-asymptotic)  approach was recently taken in \cite{BisiZygouras}, unifying 
 classes of boundary conditions at finite time through classes of Whittaker functions,  including the one considered in the present paper.  The 
 classification of stochastic particle systems with zero-range interaction \cite{Povolotsky} induces 
 integrability conditions on classes of directed-polymer models in random environment \cite{TLDSquare}, 
  including the beta and inverse-beta polymers. These models are  related by a web of dualities,
 and can exhibit features including anisotropy, degeneration to the log-gamma model in certain limits,
 and convergence of moments. They could provide natural ground for applications of the techniques
 developed in the present work. \\

 The full string expansion of the generating function  is arranged into a Fredholm Pfaffian, using 
 identities that  reproduce the algebraic structure of the derivation in the continuum 
 polymer model \cite{CalabreseLeDoussalLong}, once the appropriate kernels are known for the 
 discrete log-gamma model. The large-time
 limit of this Fredholm Pfaffian  yields the characterisation of the cumulative distribution function of the 
 (centered and rescaled) free energy of the log-gamma polymer with one free end, in terms 
 of the Tracy--Widom distribution:
\begin{equation}
 \lim_{t\to\infty} \proba\left(    \frac{ \left(  \log Z_t +  \psi\left(\frac{\gamma}{2}\right) t \right) }{  \left(-\frac{t}{8}\psi''\left(\frac{\gamma}{2}\right)  \right)^{\frac{1}{3}} }< s \right) =  \sum_{n\geq 0} \frac{1}{n!} z(n,s),
 \label{conclusion}
\end{equation}
 where $\psi$ is the digamma function. The integer parameter $n$ is the number of elements in the string expansion,
 and for each value of $n$ the $n$-string contribution reads
\begin{equation}
z(n,s) = (-1)^n\int_{{\mathbf{R}}^n}\left( \prod_{k=1}^n dx_k \right) \det[ B_s( x_i, x_j) ]_{n\times n},
\end{equation}
 and the kernel $B_s$ is defined in terms of the Airy function as
 \begin{equation}
B_s( x,y ) = \theta( x ) \Ai( x+y + s) \theta(y),
 \label{kernel}
 \end{equation}
 where $\theta( x) = \mathbf{1}( x>0)$, and the r.h.s. of Eq. \ref{conclusion} is
 the  Gaussian Orthogonal Ensemble (GOE) Tracy--Widom distribution,  denoted by $F_1(s)$.
 This universal function also governs the distribution function of the large-time limit of the
 continuum polymer model with one free end, mapped to the KPZ equation with flat boundary conditions.



\section{Review and quantities of interest}
\subsection{Definition of the model}
 
 Given a rectangular lattice with vertices labeled by the discrete values of the coordinates $x$ and $t$, 
 random Boltzmann weights denoted by $w_{x,t}$ attached to each vertex.
 In the log-gamma directed polymer model \cite{logGammaDef},  these  weights (at finite temperature set to unity)
  are  independent identically-distributed
 according to the distribution: 
\begin{equation}\label{Pgamma}
 P_\gamma(w) dw = \frac{1}{\Gamma( \gamma )} w^{-1-\gamma} e^{-1/w} dw, 
\end{equation}
 with fixed parameter $\gamma > 0$.\\

 The discrete one-dimensional log-gamma directed polymer is decribed
 with half-integer space coordinate $x$ and integer time coordinate  $t$, starting 
 at the fixed end  $(x=0, t=0)$, follows paths  with the  constraint 
 allowing only jumps from $(x,t)$ are to $\left(x \pm\frac{1}{2}, t+1\right)$.
  The  allowed values of the coordinate $x$ at even (resp. odd) times are therefore integer (resp. half-odd) numbers. The
 allowed paths are therefore equivalent to up-right directed paths on a two-dimensional lattice with integer 
  coordinates $i,j$, with starting point at $(i=1,j=1)$ and the relations $t=i+j-2$ and $x=(i-j)/2$.\\
  
Boltzmann weights are multiplicative along paths: a random quantity 
 can be associated to a path on the lattice by taking the product of the random weights along the path. 
  At time $t$, the partition function $Z_t$ of the directed polymer starting with a specified starting point, say at  $x=0$, and one free end, is defined as the sum  of the Boltzmann weights of the  paths $\phi$ with $t$ time steps and the specified starting point:
\begin{equation}
Z_t := \sum_{x\in \{0,\dots, L-1 \}}\sum_{\phi: (0,0) \rightarrow (x,t) } \prod_{(x',t')\in \phi} w_{x',t'},
\end{equation}
 where $L$ is the number of sites in one dimension. Moreover, at any value of time,
  the partition function of the polymer with one free end 
  is the sum  of the partition function $Z_t^{fixed}$ of the polymer with 
 two fixed ends (one at $0$, one at $x$), over the possible values of the final position:\\
\begin{equation}
  Z_t^{fixed}(x):= \sum_{\phi: (0,0) \rightarrow (x,t) } \prod_{(x',t')\in \phi} w_{x',t'},
 \label{fixedToFree}
\end{equation}
   which was studied in \cite{Corwin,Borodin,logGammaTLD}.\\

\subsection{Generating function }
With the aim of characterising the probability distribution function of the free energy of the log-gamma polymer
 with one free end, let us consider the $\lambda$-dependent generating function\footnote{We will sometimes drop the index $t$ in notations involving the  the partition sum.}:
\begin{equation}
g_\lambda( s ) = \overline{ \exp\left( - e^{-\lambda s} Z_t  \right)},
\label{generatingFunction}
\end{equation}
 where the overline denotes the average over the disorder introduced by the 
 random nature of the Boltzmann weights, described by Eq. \ref{Pgamma}.
 With the definition of the rescaled free energy $f$ 
\begin{equation}
 -\log Z_t =: \lambda f,
 \label{fDefinition}
\end{equation}
 where the value of $\lambda$  is assumed to scale as a growing function of time
 so that the large-time limit of the probability distribution of $f$ exists (and this  limit
 can be captured by the large-$\lambda$ limit,  see Section 4), the generating function becomes 
\begin{equation}
g_\lambda( s ) = \overline{ \exp\left(-  e^{-\lambda( s + f )}  \right)},
\end{equation}
so that the large-$\lambda$ limit of the expression to be averaged 
 under the disorder is just a step function, and  the large-$\lambda$ 
 limit of the generating function is related to the cumulative distribution function
 of $f$:\\
 \begin{equation}
 \lim_{\lambda\to\infty} g_\lambda( s) = \overline{ \mathbf{1}( f + s > 0 )} = \int_{-s}^{+\infty} P( f ) df,
\end{equation}
from which we see that the  probability distribution function of $f$ can be computed from 
 the large-$\lambda$ of the generating function.\\

 Expanding the generating function in powers of $Z_t$ in the definition of Eq. \ref{generatingFunction},
 and taking the average over disorder of each term yields the expression 
\begin{equation}
 g_\lambda^{mom}( s) := 1 + \sum_{n\geq 1} \frac{\overline{Z_t^n}}{n!} (-1)^n e^{-\lambda n s},
\label{gmom}
\end{equation}
  from which we see that computing the moments
 $\overline{Z_t^n}$ at all orders $n$ is enough to characterise $g_\lambda^{mom}$. Moreover, the parameter $\lambda$ can be thought 
 of as a chemical potential associated to the number of particles $n$. However, the moments of the log-gamma model 
 suffer from divergences at sufficiently  large orders, so that the expression of $g_\lambda^{mom}$  is only formal approach to 
 the calculation of the generating function.\\

\subsection{Divergences and analytic continuation of the moments}
 The partition function $Z_0$ is the same 
 as in the case of the polymer with fixed ends, as both models admit just one point-like path at time $t=0$.
 Its moment  of order $n$  is expressed as
 \begin{equation}
\overline{Z_0^n} =  \overline{w^n} =  \frac{1}{\Gamma( \gamma )} \int_0^\infty w^{-1 + n - \gamma} e^{-1/w} dw = \frac{\Gamma( \gamma - n ) }{ \Gamma( \gamma ) }.
\end{equation}
  This moment  does not exist as an integral for $n$ larger than the parameter $\gamma$. 
   However, its  expression can be analytically continued  to all complex values $n$ (except 
integers large enough to give rise to poles in the Gamma function). 
 Following Appendix A of \cite{logGammaTLD}, the generating function can therefore 
 be conjectured to be obtainable from formal sums of  moments  by an analytic-continuation prescription. The Mellin representation 
 of the exponential function, which can be rewritten using the Euler 
 reflection formula  as follows: 
\begin{equation}
 e^{-z} = \int_{-a + i{\mathbf{R}}} \frac{dm}{2i\pi} \Gamma( -m ) z^m = -  \int_{-a + i{\mathbf{R}}} \frac{dm}{2i \sin(\pi m)} \frac{1}{\Gamma( 1 +  m ) } z^m, 
 \label{analyticCont}
\end{equation}
where $a>0$,
allows one to express the generating function as an integral:
\begin{equation}\label{prescription}
\begin{split}
g_\lambda( s )& = \overline{ \exp\left(  e^{-\lambda s} Z  \right) }\\
&= - \int P(Z)  \left( 
\int_{-a + i{\mathbf{R}}} \frac{dm}{2i \sin(\pi m)} \frac{1}{\Gamma( 1 + m ) } e^{-\lambda s m} Z^m\right) dZ\\
&=  \int_{-a + i{\mathbf{R}}} \frac{dm}{2i \sin(\pi m)} \frac{1}{\Gamma( 1 + m ) } e^{-\lambda s m} \overline{Z^m}.\\
\end{split}
\end{equation}
  After obtaining expressions for integer moments, formal 
 sums such as $g_\lambda^{mom}$ will eventually be related to the generating function using continuation to complex values of $m$  with the 
 integration prescription of Eq. \ref{prescription}.\\

\subsection{Time-evolution of moments}

The discrete-time evolution of each path induces the following time-evolution equation
 for the partition function with fixed end:\\
\begin{equation}
 Z^{fixed}_{t+1}(x) = w_{x,t+1}\left( Z^{fixed}_t\left( x-\frac{1}{2}\right)  +  Z^{fixed}_t\left( x+\frac{1}{2}\right)\right),
 \label{zEvol}
\end{equation}
 and boundary condition $Z^{fixed}_{0}(x) = w_{0,0}\delta_{x,0}$. Let us review the analogy 
  worked out in Section 3 of  \cite{logGammaTLD} between this problem and the Schr\"odinger equation (with the Lieb--Liniger Hamiltonian \cite{LL}) that is known to govern the time-evolution of the moments in the continuum case.\\

At a given order $n$,  if the role of the wave function at time $t$ in the discrete setting of the log-gamma
 polymer is played by $\psi_t(x_1,\dots, x_n)$ defined in terms of the moments of the fixed-end partition sum by 
\begin{equation}
 \overline{\prod_{i=1}^n Z^{fixed}_t(x_i) } =: 2^{nt}\left(\frac{\bar{c}}{4}\right)^{n(t+1)}\psi_t(x_1,\dots,x_n),
 \label{timeEvolution}
\end{equation}
 and 
\begin{equation}
 \cbar = \frac{4}{\gamma-1},
\label{cbarDef}
\end{equation}
the time-evolution equation of $Z_t^{fixed}$ induces the following linear time-evolution equation,  which motivates the 
 definition of the operator $H_n$ (which is a discrete analogue of the Lieb--Liniger Hamiltonian of the quantum mechanics of bosons
 on a line), which we quote from \cite{logGammaTLD}:
 \begin{equation}
 \psi_{t+1}( x_1,\dots, x_n) = \left( \frac{1}{2^n} a_{x_1,\dots, x_n}\right) \sum_{(\delta_1,\dots,\delta_n) \in \left\{-\frac{1}{2}, \frac{1}{2}\right\}^n} \psi_t( x_1 -\delta_1,\dots, x_n - \delta_n) =: \left( H_n \psi_t\right)( x_1,\dots, x_n) ,
\label{timeEvol}
 \end{equation}
 with the notation
\begin{equation}
a_{x_1,\dots, x_n} = (\gamma-1)^n\overline{\prod_{\alpha=1}^n w_{x_\alpha, t+1}} = \prod_{x} h_{\sum_{\alpha = 1 }^n \delta_{x,x_\alpha}},\;\;\; {\mathrm{and}}\;\;\;h_n = \prod_{k=0}^{n-1}\frac{4}{4 - k \bar{c}}.
\end{equation}

%
%
Consider an orthogonal  set of normalisable eigenstates, denoted by $|\mu\rangle$  (labeled by an index $\mu$),
 of the time-independent Schr\"odinger equation associated to the operator $H_n$ for a fixed number $n$  of points, with associated wave functions $\Psi_\mu$,  
 corresponding to the eigenvalues $\theta_\mu$. This set is assumed to be complete\footnote{This set will eventually be taken to be
 indexed by the systems of strings worked out in \cite{logGammaTLD} in the thermodynamic limit, together with the eigenvalues,  and shown to be orthogonal with respect to a weighted scalar product 
 depending on the log-gamma distribution, whose explicit expression (given in Section 4.1 of \cite{logGammaTLD}) does not need to be reproduced here, as we will only use the orthogonality property rather than the explicit form of the product.}, so that 
for all values of the index $\mu$ labeling the set of states, 
 \begin{equation}
  H_n \Psi_\mu(x_1,\dots, x_n) = \theta_\mu \Psi_\mu (x_1,\dots, x_n).
\label{Schroedinger}
 \end{equation}
  From the definition of the partition function of the free-end model (Eq. \ref{fixedToFree}), we can 
 see that the moment of order $n$ of the 
 partition function $Z_t$ of the  model with one free end is given by a multiple 
sum (over the coordinates of the fixed end) of the moment of order $n$
 the partition   function with fixed ends:
\begin{equation}
 \overline{ Z^n_t} = \sum_{(x_1,\dots, x_n) \in \{0,\dots, L-1 \}} \overline{\prod_{i=1}^n Z^{fixed}_t(x_i) }.
\end{equation}  
 The time-evolution of the moment $\overline{ Z^n_t}$ can therefore be worked out using 
 the eigenvalues of the Hamiltonian $H_n$, and projecting the 
 initial value of the moment of the fixed-end model onto the eigenstates of the Hamiltonian (using the completeness assumption):\\
\begin{equation}
|\psi_0\rangle = \sum_\mu\frac{\langle \mu| \psi_0\rangle}{\langle \mu | \mu \rangle} |\mu\rangle, \;\;\; \;\;\; | \psi_t \rangle = \sum_\mu \theta_\mu^t \frac{\langle \mu| \psi_0\rangle}{\langle \mu | \mu \rangle} |\mu\rangle.
\end{equation}
 Using Eq. \ref{timeEvolution}  yields
\begin{equation}
 \overline{\prod_{i=1}^n Z^{fixed}_t(x_i)} =2^{nt}\left(\frac{\bar{c}}{4}\right)^{n(t+1)} \sum_\mu \theta_\mu^t \frac{\langle \mu| \psi_0\rangle}{\langle \mu | \mu \rangle}  \Psi_\mu (x_1,\dots, x_n),
\end{equation}
 which was calculated in \cite{logGammaTLD}. Upon permuting the sum over states and the sum over coordinates of the free ends of the polymers,
 we obtain
\begin{equation}
 \overline{ Z^n_t} =2^{nt}\left(\frac{\bar{c}}{4}\right)^{n(t+1)} \sum_\mu \theta_\mu^t \frac{\langle \mu| \psi_0\rangle}{\langle \mu | \mu \rangle}  \left( \sum_{x_1,\dots, x_n \in \{0,\dots, L-1 \}}\Psi_\mu (x_1,\dots, x_n)\right),
 \label{momentOrdern}
\end{equation}
 from which we observe that the only quantities left to compute once the solution to the fixed-end 
 problem is known are the overlaps 
\begin{equation}
 \Omega_\mu := \sum_{x_1,\dots, x_n \in \{0,\dots, L-1 \}}\Psi_\mu (x_1,\dots, x_n)
\end{equation}
 of the eigenstates with the uniform function over the discretized $n$-dimensional
 space, which are the discrete analogs of the integrals worked out in \cite{CalabreseLeDoussalLong} and
 reobtained in \cite{CalabreseLeDoussalQuench} through a symmetry argument to solve the KPZ problem with flat boundary conditions.
  We therefore have to review the set of eigenstates labeled by $\mu$  used in \cite{logGammaTLD} to solve the
 fixed-end log-gamma model in the thermodynamic limit.

\subsection{String states  in the thermodynamic limit}
 
 Eigenfunctions of the operator $H_n$ can be expressed  by coordinate Bethe Ansatz techniques
 discovered by Brunet. They are  symmetrized superpositions of plane waves,
   corresponding to the rapidities $(\lambda_\alpha)_{ 1 \leq \alpha \leq n}$:
\begin{equation}
 \Psi_\mu(x_1,\dots,x_n)= \sum _{\sigma \in \mathcal{S}_n} A_\sigma \prod_{i=1}^n e^{i\lambda_{\sigma(\alpha)} x_\alpha},
 \label{BetheBrunet}
\end{equation}
where
\begin{equation}
 A_\sigma = \prod_{1 \leq \alpha < \beta \leq n } \left(   1  +\frac{\bar{c}}{2}\frac{\signum( x_\beta - x_\alpha + 0^+)}{t_{\sigma(\alpha)} - t_{\sigma(\beta)}} \right),
\label{Asigma}
\end{equation}
 with the  notation
\begin{equation}
 t_\alpha = i \tan\left( \frac{\lambda_\alpha}{2}\right),
\label{oddMapping}
\end{equation}
  which implies that the plane waves can be expressed in terms of the family of parameters $(t_\alpha)_{ 1 \leq \alpha \leq n}$:
\begin{equation}
 z_\alpha = e^{i\lambda_\alpha} = \frac{ 1 + t_\alpha}{ 1 - t_\alpha}.
\label{ttoz}
\end{equation}
 The rapidities are names by analogy with the usual coordinate Bethe Ansatz \cite{GaudinTraduction},
 much of the structure of which can be recognised in  Eq. \ref{BetheBrunet}, up to the 
 fact that the role played in the combinatorial factors by $\lambda_\alpha$  in the Bethe Ansatz is played by $t_\alpha$
 (up to a factor of $2i$) in the Ansatz of Eq. \ref{BetheBrunet}.\\

  In the special case  $n=2$, working out the eigenvalues $\theta_\mu$ in the 
 diagonalisation problem of  Eq. \ref{Schroedinger} yields $\theta_{\mu,(n=2)} = 
\frac{1}{4}\sum_{\alpha,\beta \in \left\{ -\frac{1}{2},\frac{1}{2}\right\}} z_1^\alpha z_2^\beta = \frac{1}{4}\prod_{\alpha=1}^2(z_\alpha^{\frac{1}{2}} + z_\alpha^{-\frac{1}{2}})$, which generalises at higher orders,  yielding an  
 expression of the eigenvalues in terms of the families of parameters $(t_\alpha)_{1\leq \alpha \leq n}$ only:
\begin{equation}
 \theta_\mu =\prod_{\alpha=1}^n \frac{(z_\alpha^{-\frac{1}{2}} + z_\alpha^{-\frac{1}{2}})}{2}= \left(\prod_{\alpha=1}^n\frac{1}{1-t_\alpha^2} \right)^{\frac{1}{2}}
\label{thetaMuExpr}
 \end{equation}

Going back to the expression of the moment $\overline{Z_t^n}$ in Eq. \ref{momentOrdern}, let us use again the fact that at time
 $t=0$ the allowed paths are just points, and are  identical to the ones contributing to the partition function in the model with 
 two fixed ends. If $\mu$ denotes a system of strings whose numbers of particles sum to $n$, 
 the quantity $\langle \mu | \psi_0\rangle $ is therefore the combinatorial factor already 
 calculated in \cite{logGammaTLD} and reads
\begin{equation}
 \langle \mu | \psi_0\rangle = n!.
\label{muPsiZero}
\end{equation}

Imposing $L$-periodic boundary conditions induces the following equations for the family of rapidities: 
\begin{equation}
e^{i\lambda_\alpha L}  = \prod_{\beta\neq \alpha} \frac{ 2t_\alpha - 2t_\beta +\bar{c} }{2t_\beta - 2t_\alpha -\bar{c} },\;\;\;\;\; \;\;1\leq \alpha \leq n
 \label{BetheEquations}
\end{equation}
which are related to the Bethe equations of the continuum Lieb--Liniger model by examining the 
 role played by the parameters $t_\alpha$ in the Ansatz (Eq.  \ref{Asigma}).\\

Moreover, complex solutions to these equations  in the thermodynamic limit (of a large number $L$ of sites), when expressed in terms of the
 tangents of the rapidities (and not in terms of the rapidities themselves as in the 
 Lieb--Liniger model) 
 are arranged in {\emph{strings}} in the complex plane. Indeed, if the rapidity $\lambda_\alpha$ 
 has a non-zero imaginary part, the l.h.s. of Eq. \ref{BetheEquations} goes to zero exponentially as a 
 function of $L$, which implies that one of the factors in the r.h.s. must be zero in the thermodynamic limit, in the 
 numerator (resp. in the denominator) if the imaginary part is positive (resp. negative). This implies that there exists 
 an index $\beta$ such that $2t_\beta = 2t_\alpha \pm \cbar/2$ (with the optional sign identical to the sign of the 
  imaginary part of $\lambda_\alpha$), up to corrections vanishing exponentially\footnote{All the results of this paper
 are derived in the thermodynamic limit, neglecting corrections to the rapidities depending on the 
 size of the sysem.} at large $L$. Iterating this procedure yields
 a string with $m_j$ particles is specified by $m_j$ values corresponding to the 
 family of parameters
\begin{equation}
 t_\alpha = t_{j,a} = i\frac{k_j}{2} + \frac{\bar{c}}{4}( m_j + 1 - 2a ),\;\;\;a\in \{1,\dots, m_j\},
\label{talpha}
\end{equation}
 where the quantity $k_j$ corresponds at leading order (in a continuum scaling limit involving small rapidities and  large $\gamma$, and 
 worked out in Section 5 of \cite{logGammaTLD}) to the  momentum of a string of rapidities in the Lieb--Liniger model. 
 By an abuse of language, $k_j$ will sometimes be referred to as momentum in this paper (without taking the Lieb--Liniger limit).


\subsection{Expansion of the generating function in terms of strings}

 At fixed number $n$ of variables, each of the Bethe eigensates needed to calculate the 
 moment of order $n$ in the thermodynamic limit  can be organised into a certain integer number of  strings, denoted by $n_s$, 
 and the  numbers of particles 
 contained in these strings must sum to $n$, so that with the above notations
\begin{equation}
 n = \sum_{j=1}^{n_s} m_j.
\end{equation}
 On the other hand, at fixed number $n_s$ of strings, one can find Bethe  eigenstates 
 with  any integer number of variables. States with a fixed number of strings can therefore 
 contribute to an infinity terms in formal the expansion of the 
 generating function  in terms of moments. Following the strategy of \cite{CalabreseLeDoussalFlat,CalabreseLeDoussalLong,deNardis,logGammaTLD},
 we reorganise the expansion as a sum over the number of strings:
 \begin{equation}
  g_\lambda^{mom}( u)  =:  1 + \sum_{n_s\geq 1} \frac{1}{n_s !} Z^{mom}(n_s, u )    = 1 + \sum_{n\geq 1} \frac{\overline{Z_t^n}}{n!} (-1)^n e^{-\lambda n u},
 \label{stringExpansion}
\end{equation}
  whose consistency with the definition of  $g_\lambda^{mom}$ in Eq. \ref{gmom} induces the
 definition of  the quantity $Z^{mom}(n_s, u )$,  the contribution of the $n_s$ string sector to the generating 
 function.

\section{Localisation on parity-invariant string states}

\subsection{Symmetry argument from the overlap factor}

 In \cite{CalabreseLeDoussalQuench}, it was argued 
 that only parity-invariant systems of strings (systems of strings of rapidities in which no rapidity 
 can be found without its opposite to match) have non-zero overlap with the uniform wave function
 in the Lieb--Liniger model in the     thermodynamic limit. This symmetry argument
   provided a selection rule that confirmed the results
  of the detailed calculations of \cite{CalabreseLeDoussalLong}, in which the partition
 sums happened to localise on such systems of strings.
  As the mapping from rapidities to the $t_\alpha$ parameters (Eq. \ref{oddMapping}) is odd, the same localisation 
 should hold in the thermodynamic limit of the log-gamma polymer.\\  

 In the simplest possible case of  one string with two particles, we can check 
 this explicity by parametrizing the system with one real parameter $k$:\\
\begin{equation}
t_1 = ik + \frac{\cbar}{4},\;\;\;\; t_2 =ik -\frac{\cbar}{4},
\end{equation}
\begin{equation}
 z_1 = e^{ i \lambda_1}=\frac{4+\cbar + ik}{4-\cbar - ik} = (z_2^\ast)^{-1} =: \rho e^{i\theta},
\end{equation}
 with $\rho > 1$ and $\theta$ in the interval $[0,2\pi[$. 
 The overlap of this string state reads 
\begin{equation}
 \sum_{x_1,x_2\in\{0,\dots, L-1\}} \Psi_{k,m=2} = 2 \left( 2\sum_{x_1 = 0}^{L-1} \sum_{x_2 = x_1}^L{( \rho^{-1})}^{x_2-x_1} (e^{i\theta})^{x_1+x_2} +  \sum_{x_1 = 0}^{L-1}  e^{2i\theta{x_1}}\right).
\end{equation}
 Geometric sums allow to express at finite $L$ (in the case $\theta\neq 0)$ ratios of the form
\begin{equation}
  \frac{1}{L} \sum_{x=0}^{L} e^{ ix\theta}  = \frac{1}{L} \frac{1-e^{i(L+1)x\theta}}{1-e^{ix\theta}} 
= \frac{e^{i\frac{Lx\theta}{2}}}{L}\frac{\sin\left( ( L+1)x\theta/2\right)}{\sin\left( x\theta/2\right)},
\end{equation}
 which goes to zero when $L$ goes to infinity. On the other hand, if $\theta=0$ the l.h.s
 of the above equation is identically equal to 1, hence
\begin{equation}
 \lim_{L\to\infty}\left( \frac{1}{L} \sum_{x=0}^{L} e^{ ix\theta}\right) = \delta_{e^{i\theta},1}.
\end{equation}
 This identity implies that 
 the contribution of the string goes to zero in the thermodynamic limit if $z_1$ is not real, which condition 
 is equivalent to $k=0$. See Section 4 for the calculation of the overlap in the case of real parameters $t_1$ and $t_2$ (corresponding to $k=0$), that contributes to the $\overline{Z_t^2}$.\\

 We have to   express the contribution to the generating function of 
 systems of perfect strings consisting of $M$ strings with real  parameters $t_\alpha$, described 
 using $M$ integers $m_1,\dots, m_M$ corresponding to the number of  particles in each of these 
 strings, and $N$ pairs of strings, described by $N$ real parameters
$k_1,\dots,k_N$ (paired with their opposites $-k_1,\dots,-k_N$) and $N$ integers $n_1,\dots, n_N$ corresponding to the number of  particles in each of the
 elements of each of the $N$ pairs of strings. This structure is identical to the structure of the strings of
  rapidities on which the overlaps localise in the Lieb--Liniger model \cite{CalabreseLeDoussalQuench}. 

\subsection{Time-evolution factors of  parity-invariant string states}
  For an eigenstate consisting of one  string, 
 the eigenvalue of the time-evolution operator can be obtained by substituting the 
  string structure described in  Eq. \ref{talpha} into the expression of eigenvalues in Eq. \ref{thetaMuExpr}. This substition yields
\begin{equation}
  \theta_{m, k} =  \left( \prod_{a=1}^{m}\frac{1}{1-t_{a}^2} \right)^{\frac{1}{2}} = \left( \frac{2}{\bar{c}}\right)^{m} 
 \left(   \frac{\Gamma( -\frac{m}{2} + \frac{\gamma}{2} - i \frac{k}{\bar{c}} )\Gamma(  -\frac{m}{2} + \frac{\gamma}{2} + i \frac{k}{\bar{c}})}
  {\Gamma( \frac{m}{2} + \frac{\gamma}{2} - i \frac{k}{\bar{c}} )\Gamma(  \frac{m}{2} + \frac{\gamma}{2} +i \frac{k_j}{\bar{c}})} \right)^{\frac{1}{2}},
\label{GammaExpr}
\end{equation}
where the identity $\prod_{k=0}^{m-1}(a+k) = \Gamma( a+m)/\Gamma( a)$ has been used, together 
 with the relation $\gamma = 1 + 4/\bar{c}$ that follows from the definition of the charge $\cbar$ in Eq. \ref{cbarDef}.\\

In the  parity-invariant system of strings described above, the time-evolution factor is obtained 
 from products of the above expression:\\
\begin{equation}
\theta_{\mu \equiv\{ m_j, n_p, k_p \}_{1\leq j \leq M, 1\leq p \leq N}} = \prod_{p=1}^N \theta_{n_p,k_p}\theta_{n_p,-k_p}\prod_{j=1}^M \theta_{m_j, 0}.
\end{equation}
Using the expression \ref{GammaExpr} in terms of Gamma functions, we obtain:
\begin{equation}
 \theta_{\mu \equiv\{ m_j, n_p, k_p \}} = \left( \frac{2}{\bar{c}}\right)^{2\sum_{p=1}^N n_p + \sum_{j=1}^M m_j} 
 \prod_{p=1}^N
 \left(   \frac{\Gamma( -\frac{n_p}{2} + \frac{\gamma}{2} - i \frac{k_p}{\bar{c}} )\Gamma(  -\frac{n_p}{2} + \frac{\gamma}{2} + i \frac{k_p}{\bar{c}})}
  {\Gamma( \frac{n_p}{2} + \frac{\gamma}{2} - i \frac{k_p}{\bar{c}} )\Gamma(  \frac{n_p}{2} + \frac{\gamma}{2} +i \frac{k_p}{\bar{c}})} \right)^{\frac{1}{2}}
 \prod_{j=1}^M  \left(   \frac{\Gamma( -\frac{n_p}{2} + \frac{\gamma}{2} )}
  {\Gamma( \frac{n_p}{2} + \frac{\gamma}{2}  )} \right).
 \label{thetaStrings}
\end{equation}

\subsection{Norm of the parity-invariant states and overlap with the uniform wave function}
\subsubsection{Norms of the string states of the log-gamma model}

 The formula for the norm of the systems of $n_s$ strings in  the Lieb--Liniger model, conjectured by Gaudin \cite{GaudinTraduction}
 and proved by Korepin \cite{Korepin} was generalised in Section 7.5 of \cite{logGammaTLD} into a form that reduces to the Gaudin formula
  in the Lieb--Liniger limit. The squared norm of a state $|\mu\rangle$ consisting of a system of $n_s$ perfect strings with momenta $k_1,\dots,k_{n_s}$ and numbers
 of particles $m_1,\dots, m_{n_s}$, for a total number of particles denoted by $n=\sum_{i=1}^{n_s} m_i$, reads as the product of an 
  inter-string factor and an intra-string factor
\begin{equation}
\langle \mu|  \mu\rangle = n! L^{n_s} \prod_{1\leq i < j \leq n_s} \frac{4(k_i-k_j)^2 + \cbar^2(m_i+m_j)^2}{ 4 (k_i-k_j)^2 + \cbar^2(m_i - m_j)^2} \prod_{j=1}^{n_s}\left(\frac{m_j}{\cbar^{m_j-1}} \left(\sum_{a=1}^{m_j}\frac{1}{1-t^2_{j,a}} \right) \prod_{b=1}^{m_j } (1-t_{j,b}^2)\right),
 \label{normConj}
\end{equation}
 which reproduces the Lieb--Liniger result by
 formally by setting all the parameters $t_\alpha$ to zero. Let us call the second product over strings 
 in the above formula the deformation factor, and introduce the notation
\begin{equation}
 \tau_{m_j,k_j} = \sum_{a=1}^{m_j}\frac{1}{ 1-t_{j,a}^2},
 \label{tauDef}
\end{equation}
 where the mapping from the pair $(m_j,k_j)$ to the set of $m_j$ numbers $(t_{j,a})_{1\leq a \leq m_j}$, is 
 as always given by the definition of strings  in the thermodynamic limit (Eq.  \ref{talpha}). Using the notation introduced in Eq. \ref{GammaExpr} for the eigenvalue associated to a single string yields
 \begin{equation}
 \langle \mu | \mu \rangle =  \left( \left(\sum_{j=1}^{n_s} m_j\right)! \right) L^{n_s} \prod_{1\leq i < j \leq n_s} \frac{4(k_i-k_j)^2 + \cbar^2(m_i+m_j)^2}{ 4 (k_i-k_j)^2 + \cbar^2(m_i - m_j)^2} \prod_{j=1}^{n_s}\left(\frac{m_j}{\cbar^{m_j-1}} \tau_{m_j,k_j}  \theta_{m_j,k_j}^{-2}\right).
\end{equation}

 Specialising this expression to a system  of $M$ strings with zero momentum yields \\
\begin{equation}
|| \mu \equiv \{k=0, m_1,\dots,m_M \}||^2 =  \left (\left(\sum_{i=1}^M m_i \right)!\right) L^{M} \prod_{1\leq i < j \leq M} \frac{ (m_i+m_j)^2}{ (m_i - m_j)^2}
\prod_{j=1}^{M}\left(\frac{m_j}{\cbar^{m_j-1}} \tau_{m_j,0}  \theta_{m_j,0}^{-2}\right).
\end{equation}

 On the other hand, for the above-described  parity-invariant system of $N$ pairs of strings, the intra-string factor depending on the $2\sum_{j=1}^N m_j$ parameters $t_\alpha$
  depends only on the squares of these parameters, which take half as many distinct values, which is 
  reflected in our notations for each pair of strings by the fact that the quantities $\theta$ and $\tau$ entering the 
 deformation factor are  even functions of the momentum:
\begin{equation}
 \theta_{m_j,k_j} = \theta_{m_j,-k_j},\;\;\; \tau_{m_j,k_j} = \tau_{m_j,-k_j} , \;\;\;\;j=1,\dots,N.
\end{equation}
The deformation factor can therefore be expressed in terms of the  
 factors that would arise by specialising Eq. \ref{normConj} to a system of $N$ strings that give rise to  all the distinct values of the parameters $t_\alpha$, say 
 the strings  that have the momenta $k_1, \dots\, k_N$:
\begin{equation}
\begin{split}
|| \mu \equiv&\{ \{k_1,-k_1,\dots,k_N,-k_N\}, \{n_1, n_1,\dots, n_N, n_N\}\} ||^2 = \left( \left(2\sum_{p=1}^N n_p \right) !\right) L^{2N}\\
& \times \prod_{1\leq p < q < N}\left( \frac{4(k_p-k_q)^2 + (n_p+n_q) \cbar^2}{4(k_p-k_q)^2 + (n_p-n_q) \cbar^2}\;\frac{4(k_p+k_q)^2 + (n_p+n_q) \cbar^2}{4(k_p+k_q)^2 + (n_p-n_q) \cbar^2} \right) \prod_{p=1}^N \frac{4k_p^2 + n_p^2\cbar^2}{4k_p^2} \\
&  \;\times \prod_{p=1}^N \left(  \left( \frac{n_p}{ \cbar^{n_p-1}}\right)^2 \tau_{n_p,k_p}^2 \theta_{n_p, k_p}^{-4}\right).
\end{split}
\label{squaredNormPair}
 \end{equation}

 The squared norm of the most general relevant set of strings $|\mu\rangle$  can therefore  be written, up 
 to a combinatorial factor involving the total number of particles, as the product of the two above squared norms  by the missing inter-string 
 factors:\\
\begin{equation}
\begin{split}
\langle \mu | \mu \rangle  & =  \left( \left( 2\sum_{p=1}^N n_p + \sum_{j=1}^M m_j \right) !\right) L^{2N+M} \prod_{1\leq i < j \leq M} \frac{ (m_i+m_j)^2}{ (m_i - m_j)^2}\prod_{p=1}^N \frac{4k_p^2 + n_p^2\cbar^2}{4k_p^2} \\
& \times \prod_{1\leq p < q < N}\left(\frac{4(k_p-k_q)^2 + (n_p+n_q) \cbar^2}{4(k_p-k_q)^2 + (n_p-n_q) \cbar^2}\;\frac{4(k_p+k_q)^2 + (n_p+n_q) \cbar^2}{4(k_p+k_q)^2 + (n_p-n_q) \cbar^2}\right)\\
 &\times \prod_{1\leq p  < N, 1\leq j \leq M}\frac{4 k_p^2 + (n_p+m_j) \cbar^2}{4 k_p^2 + (n_p-m_j) \cbar^2}\\
& \times   \prod_{j=1}^{M}\left(\frac{m_j}{\cbar^{m_j-1}} \tau_{m_j,0}  \theta_{m_j,0}^{-2}\right) 
\prod_{p=1}^N \left(  \left(\frac{n_p}{ \cbar^{n_p-1}}\right)^2 \tau_{n_p,k_p}^2 \theta_{n_p, k_p}^{-4} \right)
\end{split}
\end{equation}

\subsubsection{Form of the overlaps}

In the case of a parity-invariant system of strings in the Lieb--Liniger model, 
 consisting of a pair of strings  with opposite momenta, it was argued in \cite{CalabreseLeDoussalQuench,deNardis},
 that the overlap with the uniform wave function in the Lieb--Liniger model depends on the 
 squared norm of just one string of particles (all with the same momentum), and each  parity invariant 
 single strings with zero momentum was related to the zero-momentum limit of such a pair of strings. We conjecture that the same relation 
 holds in the log-gamma model, up to deformations ensuring the factors of $\tau$ 
 compensate the contribution  of measure-theoretic factors after integration over phase space (see Subsection 3.4), so that the squared norms of the string states in the log-gamma model  (Eq. \ref{squaredNormPair})
 induce the overlap\\
\begin{equation}
\begin{split}
\Omega_\mu &= \frac{n! (L\cbar)^{M+N}}{\cbar^n} \frac{2^n}{2^M} \left( \prod_{j=1}^M\frac{m_j}{m_j!}     \tau_{m_j,0}  \theta_{m_j,0}^{-1}         \right) 
  \prod_{1\leq i<j\leq M}(-1)^{{\mathrm{min}}(m_i,m_j)}{\mathrm{sgn}}(m_i-m_j) \frac{m_i + m_j}{|m_i - m_j|}\\
&\left( \times \prod_{p=1}^N (-1)^{n_p} n_p  \tau_{n_p,k_p} \theta_{n_p, k_p}^{-2} \prod_{q=0}^{n_p -1} \frac{1}{ 4 k_p^2/\cbar^2 + q^2} \right)
\prod_{1\leq p  < N, 1\leq j \leq M}\frac{4 k_p^2 + (n_p+m_j) \cbar^2}{4 k_p^2 + (n_p-m_j) \cbar^2}\\
&\times \prod_{1\leq p < q < N}\left(\frac{4(k_p-k_q)^2 + (n_p+n_q) \cbar^2)}{4(k_p-k_q)^2 + (n_p-n_q) \cbar^2)}\;\frac{4(k_p+k_q)^2 + (n_p+n_q) \cbar^2}{4(k_p+k_q)^2 + (n_p-n_q) \cbar^2}\right).
\end{split}
\label{conj}
\end{equation}

As a check, consider the simplest case  needed for computation of the one-string contribution, namely that 
 of one string with $m=2$ particles: 
\begin{equation}
 t_1 = \frac{\cbar}{4},\;\;\;\; t_2 = -\frac{\cbar}{4},
\end{equation}
\begin{equation}
 z_1 = e^{ i \lambda_1}=\frac{4+\cbar}{4-\cbar} =\frac{\gamma}{\gamma-2}= e^{-i\lambda_2} = z_2^{-1}.  
\end{equation}
The combinatorial factors, corresponding to the permutations of two objects  denoted by $(12)$ and $(21)$, read
\begin{equation}
A_{(12)} = 1 + \signum( x_2 - x_1 + 0^+) = 2 \delta_{x_2>x_1} + 2 \delta_{x_2 =x_1},
\end{equation}
and
\begin{equation}
A_{(21)} = 1 - \signum( x_2 - x_1 + 0^+) = 2 \delta_{x_2 < x_1},
\end{equation}
so that 
\begin{equation}
\begin{split}
 \Psi_{m=2} (x_1,x_2)&= 2e^{i(\lambda_1 x_1 + \lambda_2 x_2)} \delta_{x_2 > x_1} + 2 \delta_{x_2 =x_1} + 2 e^{i(\lambda_1 x_2 + \lambda_2 x_1)} \delta_{x_2 < x_1}\\
 &=  2 z_1^{x_1-x_2}  \delta_{ x_2 - x_1 > 0 } + 2 \delta_{x_2 =x_1} + 2 z_1^{x_2-x_1}  \delta_{ x_2 - x_1 < 0 },
\end{split}
\end{equation}
and the overlap reads, introducing the variable $y=x_2-x_1$ to perform the sums on the lattice:
\begin{equation}
\begin{split}
\sum_{x_1 = 0}^{L-1}\sum_{x_2 = 0}^{L-1}\Psi_{m=2} (x_1,x_2) & = \sum_{x_1 = 0}^{L-1} \Psi_{m=2} (x_1,x_1) + 
   \sum_{x_1 = 0}^{L-1}\sum_{ -x_1 \leq y \leq L-1 -x_1,\; y\neq 0} \Psi_{m=2} (x_1, y + x_1 )\\
 & =  2L + 2\sum_{x_1 = 0}^{L-2} \sum_{1  \leq y \leq  L-1 -x_1}z_1^{-y} + 2\sum_{x_1 = 1}^{L-1} \sum_{ -x_1 \leq  y \leq -1 }z_1^{y} \\
& =  2L + 2 z_1^{-1} \sum_{x_1 = 0}^{L-2}\frac{1 - z_1^{-(L-1-x_1)}}{1-z_1^{-1}}+ 2z_1^{-1}\sum_{x_1 = 1}^{L-1}\frac{1 - z_1^{-x_1}}{1-z_1^{-1}}\\
& = 2L + 4(L-1)\frac{z_1^{-1}}{1-z_1^{-1}} - 2 \frac{z_1^{-L}}{1-z_1^{-1}} \sum_{x_1 = 0}^{L-2} z_1^{x_1} - 2 \frac{z_1^{-1}}{1-z_1^{-1}}\sum_{x_1 = 1}^{L-1} z_1^{-x_1}\\
& = 2L + 4(L-1)\frac{z_1^{-1}}{1-z_1^{-1}} - 2 \frac{z_1^{-L}}{1-z_1^{-1}} \frac{1 - z_1^{L-1}}{1-z_1} - 2 \frac{z_1^{-1}}{1-z_1^{-1}}\frac{z_1^{-1}(1-z_1^{-(L-1)})}{1-z_1^{-1}}.
\end{split}
\label{checkTwoParticles}
\end{equation}
 The dominant term in this expression is proportional to $L$, and as $z_1>1$ the partial sums of geometric series 
 converge, so that at large $L$ we obtain
\begin{equation}
\begin{split}
\sum_{x_1 = 0}^{L-1}\sum_{x_2 = 0}^{L-1}\Psi_{m=2} (x_1,x_2) & =  2(\gamma-1) L  - \left( \gamma - 2 \right)\left(\frac{3 \gamma}{2} - 1 \right) + z_1^{-L}\\
 &= \frac{8L}{\cbar} \left( 1 +   \frac{1}{8\cbar L}\left( \cbar - 4 \right)\left(\cbar + 6  \right)+O(L^{-1}z_1^{-L} ) \right).
\end{split}
\label{overlapLimit}
\end{equation}
On the other hand, specialising  Eq. \ref{conj} to the case $M=1, m_1 = 2, N=0$ yields a total number of particles $n=2$,
 and the conjectured expression for the thermodynamic limit of the overlap
\begin{equation}
  \Phi_{m=2,k=0} = \frac{4 (L\cbar)}{\cbar^2} \tau_{2,0}\theta_{2,0}^{-1},
\end{equation}
 which coincides with the large-$L$ limit of the overlap (Eq. \ref{overlapLimit}), as
 \begin{equation}
   \tau_{2,0} = \frac{2}{1-t_1^2}\ \;\;\;{\mathrm{and}}\;\;\; \theta_{2,0}^{-1} = \left( \frac{1}{(1-t_1^2)^2} \right)^{-\frac{1}{2}}.
\end{equation}


\subsection{Measure-theoretic factors}
  The sum over Bethe eigenstates in Eq. \ref{momentOrdern} involves an integration over 
 the space of rapidities. The expression of the  moment $\overline{Z_t^n}$ as integral over the 
 parameters $(k_j)_{1\leq j \leq n_s}$  therefore contains 
 a Jacobian factor induced by the string structure of the eigenstates, and  Eq. \ref{oddMapping} relating the string states to rapidities.
  This factor has been worked out 
 in \cite{logGammaTLD} in the case of the log-gamma polymer with fixed end, which does not involve the constraint 
  of parity invariance on string states. In the case of free ends, the constraints are implemented by $\delta$-measure
 factors of two kinds, with arguments containing either a single linear momentum or a pair of linear momenta, 
 and appeared in  \cite{CalabreseLeDoussalLong}. We therefore have to work out the Jacobian factors 
  in the log-gamma case, in which the parameters $k$ of strings are not identical to sums of rapidities. 
 
\subsubsection{Pairs of strings with identical number of particles and opposite linear momenta} 
Consider a parity-invariant pair of strings. They have the same number $m$ of particles
 and can be denoted as follows:
\begin{equation}
\begin{split}
  t_{1,a}& = i\frac{k_1}{2}+\frac{\cbar}{4}(m+1-2a),\;\;\;a\in \{1,\dots, m\},\\
   t_{2,a}& = i\frac{k_2}{2}+\frac{\cbar}{4}(m+1-2a),\;\;\;a\in \{1,\dots, m\},\;\;\; k_2 = -k_1.\\
\end{split}
\end{equation}
 To sum over all such Bethe  eigenstates, we have to integrate over the 
 corresponding rapidities, which are also parity-invariant combinations, with the same  
 number of elements, giving rise to the measure in phase space
\begin{equation}
 \sum_{\mu, n_s = 2, m_1 = m_2 = m, k_1 + k_2 = 0}\; \longrightarrow\;\prod_{a=1}^m d\lambda_{1,a} d\lambda_{2,a}\delta\left(\sum_{a=1}^m(\lambda_{1,a} + \lambda_{2,a})\right),
\end{equation}
 so that the constraint ensures $k_1=-k_2$ in the string space. Changing variables 
 to $k_1$ and $k_2$ gives rise to Jacobian factors:
\begin{equation}
\varphi(k_1,k_2):=\sum_{a=1}^m(\lambda_{1,a} + \lambda_{2,a}) = \frac{1}{i}\log\left(\prod_{a=1}^m
 \frac{(4+ 2ik_1+\cbar( m+1-2a))(4 + 2ik_2+\cbar( m+1-2a))}{(4- 2ik_1+\cbar( m+1-2a))(4- 2ik_2+\cbar( m+1-2a))} \right).
\end{equation}
 In sums over parity-invariant states containing several parity-invariant pairs of strings,
  each of the strings contributes such a measure-theoretic factor:
\begin{equation}
\frac{\partial}{\partial k_2}\varphi(k_1,k_2)|_{k_2 = -k_1}=\sum_{a=1}^m \frac{1}{1-t_{1,a}^2} = \tau_{m,k_1},
\end{equation}
 where we used the notation introced in Eq. \ref{tauDef}.
 The localisation constraint in the rapidity space can therefore be expressed in terms of a constraint 
 in the space of momenta with coordinates $k_1$ ans $k_2$ as 
\begin{equation}
 \delta\left(   \sum_{a=1}^m( \lambda_{1,a} + \lambda_{2,a}) \right)= \frac{1}{\tau_{m,k_1}}\delta( k_1 + k_2).
\end{equation}
On the other hand, the sum over string states transforms as:
\begin{equation}
\prod_{a=1}^m d\lambda_{i,a} =\left( \sum_{a=1}^m \frac{1}{1-t_{i,a}^2} \right) dk_i = \tau_{m,k_i} dk_i,
 \label{LebesgueTransformation}
\end{equation}
which is the transformation of the Lebesgue measure 
 on phase space that was worked out in Section 7 of \cite{logGammaTLD}. But the parity-invariance constraint
 $k_1=-k_2$ implies that $\tau_{m,k_1} = \tau_{m, k_2}$. 
There is therefore a net 
measure-theoretic factor when integrating over the parameters $k_1$ and $k_2$ with the parity-invariance 
 constraint:\\
\begin{equation}
 \prod_{a=1}^m d\lambda_{1,a} d\lambda_{2,a}\delta\left(\sum_{a=1}^m(\lambda_{1,a} + \lambda_{2,a})\right)=    
dk_1 dk_2  \delta( k_1 + k_2)\sum_{a=1}^m  \frac{1}{1-t_{1,a}^2}.
\end{equation}

\subsubsection{One string with zero momentum}
Let us call $m$ the number of particles in a
string solution to the Bethe equations, with zero parameter $k$ (and drop the index $j$ labeling the string), 
 in the thermodynamic limit, at leading order in the size of the system (dropping the exponentially small
 corrections):
\begin{equation}
 t_a = i\frac{k}{2}+\frac{\cbar}{4}(m+1-2a),\;\;\;{\mathrm{with}}\;\;\;k=0,\;\;\; a\in \{1,\dots, m\}.
\label{notationOneString}
\end{equation}
 As the mapping from the rapidities  to these parameters is odd, summing over all such 
 string states involves integrating against the phase space measure of all the 
 corresponding rapidities:\\
\begin{equation}
\lambda_a = \frac{1}{i} \log\left( \frac{4 +2 ik + \bar{c} (m+1 - 2a)}{4 -2 ik + \bar{c} (m+1 - 2a)}\right).
\end{equation}
The sum of rapidities of all the particles in a string
\begin{equation}
\sum_{a=1}^m \lambda_a = \frac{1}{i} \log \prod_{a=1}^m\left( \frac{4 +2 ik + \bar{c} (m+1 - 2a)}{4 -2 ik + \bar{c} (m+1 - 2a)}\right).
\end{equation}
is  zero if and only if $k=0$, since the family of integers $\{m+1 -2a\}_{1\leq a\leq m}$ is
 parity-invariant. The measure-theoretic factor therefore yields a delta-function of the scalar parameter $k$, up to 
 a division by:
\begin{equation}
 \frac{\partial}{\partial k} \prod_{a=1}^m\left( \frac{4 +2 ik + \bar{c} (m+1 - 2a)}{4 -2 ik + \bar{c} (m+1 - 2a)}\right)|_{k=0} = 4i\sum_{a=1}^m\frac{1}{4 + \bar{c}(m+1-2a)}= \frac{i}{1 + t_a},
\end{equation}
 so that the integration measure over the phase space of the one-string parity-invariant sector reads (using again Eq. \ref{LebesgueTransformation}
 for the Lebesgue measure):
\begin{equation}
\prod_{a=1}^m d\lambda_a\delta\left(\sum_{a=1}^m \lambda_a\right)= 
  \tau_{m,0} dk \frac{1}{\left|\sum_{a=1}^m\frac{1}{1+ t_a}\right|} \delta(k).
\end{equation}
 The above expression can be related to the quantity $\tau_{m,0}$ using the 
 parity invariance of the family of $t_a$ parameters:
\begin{equation}
\sum_{a=1}^m \frac{1}{1 +t_a} = \frac{1}{2}\sum_{a=1}^m\left( \frac{1}{1 +t_a} +\frac{1}{1 -t_a}    \right) =  
\frac{1}{2} \sum_{a=1}^m \frac{2}{1 -t_a^2} = \tau_{m,0},
\end{equation}
 yielding the measure-theoretic contribution
\begin{equation}
\prod_{a=1}^m d\lambda_a\delta\left(\sum_{a=1}^m \lambda_a\right)=
\frac{\tau_{m,0}}{|  \tau_{m,0}|} \delta( k ) dk.
\label{measureOneString}
\end{equation}
 The constraint $\delta(k)$  ensures that the string has zero imaginary part.
 In sums over parity-invariant states containing several strings with zero momentum,
  each of the strings contributes such a measure-theoretic factor. These factors compensate 
 the factors from the Lebesgue measure, up to a sign, which can be fixed
 by requiring that setting all the $t_\alpha$ parameters to 0 should yield
 the Lieb--Liniger formula.

\section{The one-string contribution to the generating function at large time}

 Using the  measure worked out in Eq. \ref{measureOneString}, leaves the sum over eigenstates initially labeled by $\mu$
 as a sum over the number of particles $m$ in the string states (as in does in the continuum Lieb--Liniger model with one free-end 
in \cite{CalabreseLeDoussalLong}, up to deformation factors coming from the squared norms of the string states and from the
 overlap factors). From the definition of the string expansion of the generating function  in terms of  integer moments in Eq. \ref{stringExpansion},
  we find the formal series 
\begin{equation}
Z^{mom}( n_s=1,u ) = \sum_{m=1}^\infty \frac{(-1)^m e^{-\lambda mu}}{m!} \left( 2^{mt}\left( \frac{\bar{c}}{4}\right)^{m(t+1)} \theta_{m,0}^t \frac{m!}{\langle k=0,m| k=0, m\rangle}\Omega_{m,0}\right),
 \label{oneStringFirstGo}
\end{equation}
 where use has been made of the first term of the string expansion of  each moment $\overline{Z_t^m}$, read off from Eq. \ref{momentOrdern}, and specialised to a string of $m$ particles with $k=0$, together with the combinatorial factor of  Eq. \ref{muPsiZero}.
 Moreover, the square of the norm of the single-string state for all values of the momentum and of the number of particles 
   is obtained as a particular case of Eq. \ref{normConj}:
 \begin{equation}
\langle k=0,m| k=0, m\rangle = m! \frac{m L}{\cbar^{m-1}}\tau_{m,0}\theta_{m,0}^{-2},
\end{equation}
and the overlap is read off by specialising Eq. \ref{conj} to the case of one string with $m$ particles
\begin{equation}
 \Omega_{m,0} \sim_{L\to\infty}  \frac{2^{m-1}mL}{\cbar^{m-1}}\tau_{m,0}\theta_{m,0}^{-1}.
\end{equation}
Moreover the time-dependent contribution can be read off from Eq. \ref{thetaStrings}
\begin{equation}
\theta_{k=0,m} = \left(\frac{2}{\bar{c}}\right)^m \left( \frac{\Gamma( -\frac{m}{2} + \frac{\gamma}{2})} {\Gamma( \frac{m}{2} + \frac{\gamma}{2})}  \right).
\end{equation}

 The factors of $\sum_{a=1}^{m_j} \frac{1}{1-t_{j,a}^2}$ come in inverse powers from the 
 Lebesgue measure and delta-function on one side, and from the overlap and squared norm of the string state
 on the other side. The one-string contribution 
 is therefore given by a sum over the number of particles:\\
\begin{equation}
 Z^{mom}(n_s = 1, u) =\frac{1}{2} \sum_{m = 1}^\infty ( -1)^m 2^{m(t+1)} 
\left( \frac{\bar{c}}{4}\right)^{m(t+1)}\frac{1}{ m!} \left(\frac{2}{\bar{c}}\right)^{m(t+1)}\left( \frac{\Gamma\left( -\frac{m}{2} + \frac{\gamma}{2}\right)}{ \Gamma\left( \frac{m}{2} + \frac{\gamma}{2}\right)}\right)^{t+1} e^{-\lambda m u}.
\end{equation}
The analytic continuation to complex values of $m$ inspired by the Mellin representation of the exponential function \cite{logGammaTLD} yields
 the integral form of the one-string contribution to the denerating function:
\begin{equation}
Z( n_s=1, u ) =- \frac{1}{2}\int_{C} \frac{ds}{2i\sin( \pi s )} e^{-\lambda u s}\frac{1}{\Gamma( s + 1)} \left( \frac{\Gamma\left( -\frac{s}{2} + \frac{\gamma}{2}\right)}{ \Gamma\left( \frac{s}{2} + \frac{\gamma}{2}\right)}\right)^{t+1},
\end{equation}
 where $C = a + i\mathbf{R}$, in the notations of Eq. \ref{analyticCont}. 
  Let us define the function 
\begin{equation}
 J_\gamma(s ) := \log\left(\frac{\Gamma\left( -\frac{s}{2} + \frac{\gamma}{2}\right)}{ \Gamma\left( \frac{s}{2} + \frac{\gamma}{2} \right) }\right).
\end{equation}
  Rescaling the integration 
 variable by a factor of $\lambda$ by the change of variable $\tilde{s} = \lambda s$  should allow us use  the saddle-point 
 method in the large-$\lambda$ limit.  The one-string contribution to the generating function therefore reads
\begin{equation}
Z(n_s= 1, u ) =- \frac{1}{2} \int_{C} \frac{d\tilde{s}}{2i \lambda \sin( \pi \frac{\tilde{s}}{\lambda} )} \frac{1}{\Gamma\left(\frac{\tilde{s}}{\lambda} + 1\right)} 
\exp\left(  -{\tilde{s}} u +  t' J_\gamma\left(\frac{\tilde{s}}{\lambda} \right)  \right),
\end{equation}
 where $t'=t+1$, which in the large-time limit is equivalent to time. 
The function $J_\gamma$  can be Taylor-expanded around $s=0$ as
\begin{equation}
 J_\gamma(s ) = - \psi\left(\frac{\gamma}{2}\right)s- \frac{1}{24} \psi''\left(\frac{\gamma}{2}\right) s^3+ O( s^5),
\end{equation}
 where $\psi=\Gamma'/\Gamma$ denotes the digamma function.\\

\begin{equation}
Z( n_s=1, u ) =-\frac{1}{2}  \int_{C} \frac{d\tilde{s}}{2i \lambda \sin( \pi \frac{\tilde{s}}{\lambda} )} \frac{1}{\Gamma\left(\frac{\tilde{s}}{\lambda} + 1\right)} 
\exp\left(  
-{\tilde{s}} u -
    \psi\left(\frac{\gamma}{2}\right)\frac{\tilde{s}}{\lambda}  t'  - \frac{t'}{24 \lambda^3 } \psi''\left(\frac{\gamma}{2}\right) {\tilde{s}}^3+ O( t' \lambda^{-5})
  \right)
 \label{ZoneString}
\end{equation}
 Let us shift the energy by so that the term proportional to  $ts$ in the argument of the exponential disappears, meaning 
  that the large-$\lambda$ limit will address  the distribution function of  the random variable $\log{Z_t} + \psi(\gamma/2) t$, whose random character is entirely due to $Z_t$. 
 Let us   adjust the parameter $\lambda$  using the Airy representation of the exponential 
 of a cubic function (which in the continuum model was used 
 to tame the rapid growth of the  moments, that were defined at all orders but whose energy grows as the cube of the 
 order, \cite{CalabreseLeDoussalLong}, whereas in the log-gamma model with fixed ends it was ). The identity
\begin{equation} 
e^{\frac{w^3}{3}} = \int_{-\infty}^{\infty} dy \Ai( y ) e^{yw}
 \label{Airy}
\end{equation}
 should allow us to rewrite  the cubic term in $\tilde{s}$  as an exponent that is linear in the parameter $\lambda$. With the choice
\begin{equation}
 \lambda^3 = -\frac{1}{8}\psi''\left(\frac{\gamma}{2}\right) t,
\label{lambdaFix}
\end{equation}
  the large-time and large-$\lambda$ limits become equivalent (and one can check that replacing $t'$ with $t$ altogether in Eq. \ref{ZoneString}, with 
  time scaling as the cube of $\lambda$, yields a correction of order $O(\lambda^{-3}$, which is subdominant as $t\lambda^{-5}\sim_{\lambda\to\infty}\lambda^{-2}$). 
   Going back to the definition of the rescaled free energy denoted by $f$  (in Eq. \ref{fDefinition}), and using the notation $f$ again
 for the rescaled free energy of the free-end model with the above-described shift in energy, we  can write
\begin{equation}
 f =  \frac{-2}{  \left( \psi''\left(\frac{\gamma}{2}\right) t \right)^{\frac{1}{3}}}\left(  \log Z_t +  \psi\left(\frac{\gamma}{2}\right) t \right),
 \end{equation}
 The large-$\lambda$ limit of the one-string contribution $Z(1,u)$ of   Eq. \ref{ZoneString}, together 
 with the definition of $\lambda$ as a function of time in Eq. \ref{lambdaFix}, characterises the one-string contribution to the  cumulative 
 distribution function of the rescaled free energy $f$. The shift in the energy levels and the definition of $\lambda$ in terms of time and the 
 parameter $\gamma$ we have just made induce all the higher-order contributions to the string expansion of the  generating function.\\

 Using the Airy representation of Eq. \ref{Airy}, we obtain the one-string contribution as a double integral, in which we can 
 shift the argument of the Airy function by a change of variable: 
\begin{equation}
\begin{split}
Z( n_s=1, u ) & \underset{\lambda\to\infty}\sim \frac{1}{2}\int_{-\infty}^{+\infty} dy \Ai( y )\int_{-a + i {\mathbf{R}}} \frac{d\tilde{s}}{2i \lambda \sin( \pi \frac{\tilde{s}}{\lambda} )} \frac{1}{\Gamma\left(\frac{\tilde{s}}{\lambda} + 1\right)} 
\exp\left( (y-u) {\tilde{s}}   + O(  \lambda^{-2})
  \right)\\
& \underset{\lambda\to\infty}\sim \frac{1}{2} \int_{-\infty}^{+\infty} dy \Ai( y + u)\int_{-a + i {\mathbf{R}}} \frac{d\tilde{s}}{2i \lambda \sin( \pi \frac{\tilde{s}}{\lambda} )} \frac{1}{\Gamma\left(\frac{\tilde{s}}{\lambda} + 1\right)} 
\exp\left(  y{\tilde{s}}   + O(  \lambda^{-2})
  \right) .
\end{split}
\end{equation}
 Using the fact that the Laplace transform of the step function is expressed by the identity
\begin{equation}
\int_{C} \frac{ds}{2i\pi s} e^{ sy} = \mathbf{1}( y > 0 )
\end{equation}
  yields the large-$\lambda$ limit of the one-string contribution to the generating function as:
\begin{equation}
Z( n_s=1, u ) \underset{t\to\infty}\sim - \frac{1}{2} \int_{0}^\infty   Ai( y + u ) dy.
 \label{oneStringComb}
\end{equation}
 As in the continuum Lieb--Liniger model, assuming 
  a Fredholm determinant structure for the cumulative distribution 
 function of the rescaled  free energy, as in the right-hand-side  of 
 Eq. \ref{conclusion}, the one-string  calculation yields 
 the centering and scaling prescriptions that ensure that allow 
 to write the left-hand side in terms of the parameters of the model only.
 Moreover,  it yields the (opposite of the) trace of the kernel $B_s$, which supports the 
  form of the kernel announced in Eq. \ref{kernel}.
 Indeed, the integer parameter organising the expansion of the Fredholm
 determinant in powers of the kernel is identical to the number of strings.\\

\section{Pfaffian structure of the generating function}
 
  The odd mapping  from rapidities to  the string states of the log-gamma polymer, and the measure-theoretic 
 factors  worked out for parity-invariant string states, preserve much of the combinatoric structure of the 
 generating function of the Lieb--Liniger model, 
 based on the Pfaffian identities which we borrow \cite{CalabreseLeDoussalLong}, that allow to rewrite the inter-string products in
 in the factors $\Omega_\mu (\langle \mu|\mu\rangle)^{-1}$  terms of Pfaffians
 of antisymmetric matrices defined  as $\pfaffian( (X_i-X_j)(X_i + X_j)^{-1})$ from a vector $X$ with $n_s = 2N+M$ components containing
 linear combinations the parameters describing the strings (see 
 Eqs 111--115 of \cite{CalabreseLeDoussalLong})
  The formal string expansion in terms of integer moments  is therefore a modification 
of the formula 136 in Section 6 of \cite{CalabreseLeDoussalLong}, where the parameters $(k_i)_{1\leq i \leq n_s}$ have been rescaled by a factor of $\cbar$, which is consistent with the
 replica result based on the overlap with the parity-invariant states:
\begin{equation}
\begin{split}
Z^{mom}(n_s, u) = \sum_{ (m_1,\dots, m_{2n_s}) \in ({\mathbf{N}}^\ast)^{n_s} } \prod_{j=1}^{n_s} \int_{-\infty}^{+\infty}\frac{dk_j}{2\pi}  \prod_{q=1}^{m_j}\frac{-2}{2ik_j + q}  &\Xi(m_j, \cbar k_j)^{\frac{t+1}{2}} e^{-\lambda m_j u}\\
 &\times{\mathcal{P}}((k_i)_{1\leq i\leq n_s},  (m_i)_{1\leq i\leq n_s}),
\end{split}
\end{equation}
 where the Pfaffian factor $\mathcal{P}$ is evaluated on a matrix whose size is twice the number of strings, and the $i$-th string 
 is specified by the real number $k_i$ and the integer number of particles $m_i$, so that the two terms in the 
 upper-left block implement the selection rules of the localisation on parity invariant states:
\begin{equation}
\begin{split}
  \mathcal{P}&( (k_i)_{1\leq i\leq n_s},  (m_i)_{1\leq i\leq n_s}) = \hspace{15cm}\\
&\pfaffian \left(
   \begin{array}{c c}
     \frac{2\pi}{2ik_i}\delta(k_i + k_j) (-1)^{m_i} \delta_{m_i,m_j}+ \frac{(2\pi)^2}{4}\delta(k_i) \delta(k_j)(-1)^{{\mathrm{min}}(m_i,m_j)}{\mathrm{sgn}}(m_i-m_j) &           -\frac{1}{2} (2\pi) \delta( k_j)  \\
      \frac{1}{2} (2\pi) \delta( k_i)&   \frac{2ik_i + m_i -2i k _j - m_j}{2ik_i + m_i + 2i k_j + m_j},\\
   \end{array}
   \right)_{2n_s \times 2n_s},
\label{Pfaffian}
\end{split}
\end{equation}
and  the time-evolution is governed by the function of $m$ and $k$, which contains the parameter $\gamma$ of the 
 distribution of Boltzmann weights (even though we dropped the symbol $\gamma$ from the notation):
\begin{equation}
\Xi(m,k) = \frac{\Gamma(-\frac{m}{2} +\frac{\gamma}{2} -i k ) \Gamma( -\frac{m}{2}+ \frac{\gamma}{2} +i  k )}
{\Gamma(\frac{m}{2} +\frac{\gamma}{2}-i k )\Gamma(\frac{m}{2}+ \frac{\gamma}{2} +i k)}.
 \label{XiDef}
\end{equation}

 The compensation of the 
 factors $\tau_\mu$ in both pairs of strings and strings with zero parameter $k$, ensured by the measure-theoretic factors,
 ensures that the only modification of the Lieb--Liniger is carried by the time-dependent factors (with a shift of time 
 induced by the $\theta_\mu$ factors in the overlaps and norms of Bethe  eigenstate labeled by $\mu$). 
 The one-string constribution can be seen to reproduced the result derived above, as the constribution 
 of the Pfaffian factor just consists of the upper-right entry of the matrix, enforcing the parity-invariant constraint 
 on the string.
 Again borrowing from \cite{CalabreseLeDoussalLong}, we
  can represent each of the rational fractions in the string parameters entering the Pfaffian as a double 
 integral of an exponential function, with a constraint, starting from the identity:
\begin{equation}
\frac{2ik_i + \lambda m_i -2i k _j - \lambda m_j}{2ik_i +\lambda  m_i + 2i k_j +\lambda  m_j}= \int_0^{\infty} dv_i \int_0^\infty dv_j \delta(v_i - v_j) (\partial_{v_i}- \partial_{v_j})
e^{-v_i(2ik_i +\lambda  m_i) - v_j(2i k_j +\lambda  m_j)}.
\end{equation}
  Let us  extract the exponential integrands from the matrix by use of the following multilinearity property of the Pfaffian,
 written in terms of square matrices $A$ and $B$ of size $n_s$, and vectors $U$ and $V$ with $n_s$ components, 
\begin{equation}
{\mathrm{Pf}} \left(
   \begin{array}{c c}
       A_{ij} &      \lambda_i U_iV_j  \\
      -\lambda_j U_jV_i &   \lambda_i\lambda_j B_{ij} \\
   \end{array}
   \right)_{2n_s \times 2n_s}=\left(  \prod_{j=1}^{n_s} \lambda_j\right)
{\mathrm{Pf}}\left(
   \begin{array}{c c}
       A_{ij} &      U_iV_j  \\
      -U_jV_i &   B_{ij} \\
   \end{array}
   \right)_{2n_s \times 2n_s},
\end{equation}
 to obtain the $n_s$-string term as a multiple integral with all sums outside the Pfaffian:
\begin{equation}
\begin{split}
Z^{mom}(n_s, u) = \sum_{\{m\} \in ({\mathbf{N}}^\ast)^{n_s} } \prod_{j=1}^{n_s} \int_{v_j\geq 0 }\int_{-\infty}^{+\infty}\frac{dk_j}{2\pi}
 e^{-\lambda m_j u - v_j(2ik_j + \lambda m_j)} \prod_{q=1}^{m_j}\frac{-2}{2ik_j/\lambda + q}\Xi(m_j, k_j/\lambda)^{\frac{t+1}{2} }\\
{\mathrm{Pf}}\left(
   \begin{array}{c c}
      \frac{2\pi}{2ik_i}\delta(k_i + k_j) (-1)^{m_i}\delta_{m_i,m_j} + \frac{(2\pi)^2}{4}\delta(k_i)\delta(k_j)(-1)^{{\mathrm{min}}(m_i,m_j)}{\mathrm{sgn}}(m_i-m_j) &     \pi \delta( k_i)\delta(v_j)  \\
    -\pi \delta( k_j)\delta(v_i)     &  2 \delta'(v_i - v_j )\\
   \end{array}
   \right)_{2n_s \times 2n_s}.\\
\end{split}
\end{equation}

Putting back all the sums into the Pfaffian by the multilinearity property we 
 express the $n_s$-string term as the integral of  a Pfaffian:
\begin{equation}
g^{mom}_\lambda(s) = 1 + \sum_{n_s\geq 1} \frac{1}{n_s!} \prod_{j=1}^{n_ s}\int_{v_j\geq 0}{\mathrm{Pf}}
\left(
   \begin{array}{c c}
     \kernel^{mom}_{11}(v_i,v_j)&        \kernel^{mom}_{12}(v_i,v_j)    \\
    -  \kernel^{mom}_{12}(v_i,v_j)&   \kernel^{mom}_{22}(v_i,v_j) 
   \end{array}
\right)_{2n_s\times n_s}
\label{PfaffianMom}
\end{equation}
where the kernels entering the Pfaffian are expressed as formal sums, and 
 apart from 
\begin{equation}
\kernel^{mom}_{22}(v_i,v_j)   = 2 \delta'(v_i - v_j ),
\end{equation}
 contain sums over the numbers of particles, and time dependent powers of the kernel $\Xi$
\begin{equation}
\kernel^{mom}_{12}(v_i,v_j)   = -\pi \delta(v_j) \sum_{ m_i\geq 1}\frac{(-2)^{m_i}}{m_i!}\Xi(0, m_i)^{\frac{t+1}{2}} e^{-\lambda m_i( s + v_i)}= 
\tilde{\kernel}(v_i)\delta(v_j),
\end{equation}
which is a rank-one operator (on $L^2({\mathbf{R}})$), with 
\begin{equation}
\tilde{\kernel}(v) =  \pi\sum_{ m\geq 1} \frac{(-2)^m}{m!}\Xi(0, m)^{\frac{t+1}{2}} e^{-\lambda m( s + v)}
\label{ktilde}
\end{equation}

\begin{equation}
\kernel^{mom}_{11}(v_i,v_j)   = Q^{mom}_1(v_i,v_j) + Q^{mom}_2(v_i,v_j),
\end{equation}
 with the following two formal kernels, adapted to the two sectors on which the phase-space measure 
 of the two-string states localises (respectively the sector with one pair of strings with identical numbers 
 of particles, and the sector with two strings with independent of particles and purely imaginary rapidities):

\begin{equation}\label{Q2}
\begin{split}
 Q^{mom}_1(v_i,v_j) = \int_{k_i} \frac{\pi}{ik_i}\sum_{ m_i\geq 1}(-1)^{m_i}\Xi(m_i,k_i/\lambda)^{\frac{t+1}{2}}\Xi(m_i,-k_i/\lambda)^{\frac{t+1}{2}}
e^{-\lambda m_i(2s + v_i + v_j) -2ik_j(v_j-v_i))} \prod_{q=1}^{m_i} \frac{4}{ q^2 + 4 k_i^2/\lambda^2},
\end{split}
\end{equation}

\begin{equation}\label{Q2}
 Q^{mom}_2(v_i,v_j) =\frac{(2\pi)^2}{4} \sum_{ m_i\geq 1}\sum_{ m_j\geq 1}\frac{e^{-m_i\lambda ( s  + v_i)}}{m_i!}\frac{e^{-m_j\lambda ( s  + v_j)}}{m_j!}\Xi( m_i,0)^{\frac{t+1}{2}}\Xi(m_j,0)^{\frac{t+1}{2}}(-1)^{{\mathrm{min}}(m_i,m_j)}{\mathrm{sgn}}(m_i-m_j).
\end{equation}

   Applying the analytic continuation prescription to the above kernels should yield
  a structure  whose large-time limit can be studied, yielding  the higher-order 
 string expansion of the distribution function of the rescaled free energy. 
  Pushing the string expansion up to the second order in  $n_s$,  is high enough 
 to make all the  kernels appear.\\

\section{The two-string contribution to the generating function at large time}
In the two-string sector there are contributions from a pair of strings
 with opposite momentum parameters and identical number of particles (which require 
 the integration of the weighted localised measure on pairs of strings worked out 
 in the previous section), and contributions 
 from two strings with zero parameter $k$, and any numbers of particles. 
 The fact that $\kernel_{12}$ is a rank-one operator allows use to make use of 
 the Fredholm identities of Appendix G of \cite{CalabreseLeDoussalLong}, which yield
\begin{equation}
\begin{split}
Z( n_s=2, s ) &= -{\mathrm{Tr}}(\kernel^{mom}_{11 } \kernel^{mom}_{22} ) = \int_{v_1} \int_{v_2} \kernel^{mom}_{11}( v_1, v_2 )\kernel^{mom}_{22}(v_2, v_1 )\\
 &= -2  \int_{v_1} \int_{v_2} \kernel^{mom}_{11}( v_1, v_2 )\delta'(v_2 - v_1 )=  +2 \int   \kernel^{mom}_{10}(v,v) dv,
\end{split}
\end{equation} 
with 
\begin{equation}
 \kernel^{mom}_{10}(v_1,v_2)  = \partial_{v_1} \kernel^{mom}_{11}(v_1,v_2) = \partial_{v_1} Q^{mom}_1(v_1,v_2)  + \partial_{v_1} Q^{mom}_2(v_1,v_2) .
\end{equation} 

\subsection{One  pair of strings with identical numbers of particles}
 Let us denote by $m$ the number of particles in each strings. 
  The product of $m$ rational fractions of the $k$ parameter can be put 
 into a form more suitable to analytic continuation to complex values of $m$:\\
\begin{equation}
\begin{split}
  \prod_{ q = 1}^m\frac{1}{4k^2/\lambda^2 + q^2} &= \prod_{q=1}^m\frac{1}{(q-2ik/\lambda)( q + 2ik/\lambda)} \\
&= \frac{1}{(1-2ik/\lambda)_m(1+2ik/\lambda)_m} = 
\frac{\Gamma( 1 - 2ik/\lambda )\Gamma( 1+2ik/\lambda)}{\Gamma( 1-2ik/\lambda +m )\Gamma(1 + 2ik/\lambda + m)}.
\end{split}
\end{equation}
 where use has been been of the Pochhammer symbol. The analytic-continuation prescription of the kernel $Q^{mom}_1$
therefore yields  an expression whose  large-$\lambda$ limit  (which is equivalent to the large-time limit because of the choice of $\lambda$ as a function 
 of time we made in Eq. \ref{lambdaFix}), can yield a $\lambda$-independent equivalent, once the 
  change of variable $m' = \lambda m$ has been performed:
\begin{equation}\label{Q1}
\begin{split}
Q_1(v_1,v_2) = &  -  \int_{-a + i{\mathbf{R}}} \frac{dm}{2i\pi \sin(\pi m)} \int_{-\infty}^\infty dk_1  \int_{-\infty}^\infty dk_2 \delta( k_1+k_2)  \frac{\pi}{ik_1}\Xi(m,k_1/\lambda)^{\frac{t+1}{2}}\Xi(m,k_2/\lambda)^{\frac{t+1}{2}}\\
& e^{-\lambda m (2s + v_1 + v_2) +2i( k_1v_1 + k_2v_2)}\frac{\Gamma( 1  + 2ik_1/\lambda )\Gamma( 1+2ik_2/\lambda)}{\Gamma( 1-2ik_1/\lambda +m )\Gamma(1 + 2ik_2/\lambda + m)}\\
& = \int_{-a + i{\mathbf{R}}} \frac{dm'}{2i\pi \lambda\sin(\pi m'/\lambda )} \int_{-\infty}^\infty dk_1  \int_{-\infty}^\infty dk_2 \delta( k_1+k_2)  \frac{\pi}{ik_1}\Xi(m'/\lambda, k_1/\lambda )^{\frac{t+1}{2}}\Xi(m'/\lambda, k_2/\lambda)^{\frac{t+1}{2}}\\
& e^{-m' (2s + v_1 + v_2) +2i( k_1v_1 + k_2v_2)}\frac{\Gamma( 1  + 2ik_1/\lambda )\Gamma( 1+2ik_2/\lambda)}{\Gamma( 1-2ik_1/\lambda +m'/\lambda )\Gamma(1 + 2ik_2/\lambda + m'/\lambda)}
\end{split}
\end{equation}
To capture the large-time limit of the contribution of two strings of opposite momenta, we can again use the 
 Laplace method which involves the expansion of the logarithm of  $\Xi(m,k)$ around $(0,0)$. Since 
 this logarithm   is odd in $m$ and even in $k$ (see Eq. \ref{XiDef}), we can write it in the form:\\
\begin{equation}\label{Taylorkm}
\begin{split}
 \Phi( m, k )& : = \log \Xi(m,k)\\
 & = -2 \psi\left( \frac{\gamma}{2}\right) m  + \psi''\left( \frac{\gamma}{2}\right)  mk^2  
- \frac{1}{12}   \psi''\left( \frac{\gamma}{2}\right) m^3 + \sum_{p\geq 2} \sum_{q=0}^{p} \xi_{p,q} k^{2q}m^{2p+1 - 2q}\\
& =: -2 \psi\left( \frac{\gamma}{2}\right) m + \tilde{\Phi}( m, k ).
\end{split}
\end{equation}
where the coefficients $\xi_{p,q}$ can be expressed in terms of derivatives of the digamma function at $\gamma/2$, ensuring that 
 contributions of the same order in $k$ and $m$ will play the same role   in the large-time limit if both variables are scaled by the same power of the parameter $\lambda$ (see Eq. \ref{scaledKernel}). The linear term in $m$
 will be discarded in the expansion, as the shift in the free energy levels 
 implies that $\tilde{\Phi}$ will be substituted to $\Phi$ in all the expressions (as $\Phi(m,0)=2 J_\gamma( m )$, whose linear term in $m$ dictated the 
  centering of the free energy in the one-string contribution). Let us use a Laplace representation
in the expansion parameter $m$:
\begin{equation}
 \tilde{\Xi}(m,k)^t = \int_0^\infty \Upsilon( t, k, u) e^{-m u } du
\end{equation}
where $\Upsilon$ is an inverse Laplace transform, with the free energy shifted as prescribed in the paired-string  contribution:
\begin{equation}
\Upsilon( t, k, u)   = \int_{C} \frac{dz}{ 2i\pi}   \tilde{\Xi}(m,k)^t e^{zu}.\\
\end{equation}
 The change of variable defined by  $u = \lambda v$ yields
\begin{equation}
\tilde{\Xi}(m,k)^t = \lambda \int_0^\infty \Upsilon( t, k, \lambda v)   e^{-m\lambda v} d v =  \lambda \int_0^\infty dv   \int_{C} \frac{dz}{ 2i\pi}  \tilde{\Xi}(k,z)^t  e^{z\lambda v-m\lambda v  },
\end{equation}
and the change of variable $w = \lambda z$ puts the expression in a form suitable to a series expansion of the exponent $\tilde{\Phi}$ around zero:
\begin{equation}
 \tilde{\Xi}(m,k)^t =  \int_0^\infty dv   \int_{C} \frac{dw}{ 2i\pi} \exp\left(  t \tilde{\Phi}\left( \frac{w}{\lambda}, k \right) +  v(w-\lambda m) \right).
\end{equation}
 The argument of the  exponential function in the above expression has terms 
 of order 0 and 1 in $m$. Let us insert the expansion \ref{Taylorkm} and take care of the cubic term in $m$  using  the Airy 
  function and the scaling $\psi''(\gamma/2) t = -8 \lambda^3$ that was identified in Eq. \ref{lambdaFix} based on the one-string contribution:
\begin{equation}\label{XiSaddle}
 \tilde{\Xi}(m,k)^{\frac{t}{2}} =  \int_0^\infty dv   \int_{C} \frac{dw}{ 2i\pi} \exp\left(    v(w  - \lambda m)  -  4 \lambda^2  wk^2 + \frac{w^3}{3} 
-4\frac{\lambda^3}{\psi''\left( \frac{\gamma}{2}\right)}  \sum_{p\geq 2} \sum_{q=0}^{p} \xi_{p,q}\lambda^{2q-2p-1} k^{2q} w^{2p+1 - 2q} \right).
\end{equation}
 As we need to insert the value of the kernel  at $(k/\lambda, m'/\lambda)$ into the integrand of the expression of $Q_1$, Eq. \ref{Q1}, 
 the higher-order terms with the are all be weighted by a power of $\lambda$ equal to the total degree of the correction:
\begin{equation}\label{scaledKernel}
\begin{split}
 \tilde{\Xi}&\left(\frac{m}{\lambda},\frac{k}{\lambda}\right)^{t/2} =  \int_0^\infty dv   \int_{C} \frac{dw}{ 2i\pi} \exp\left(    vw  -  v m  - 4 wk^2  + \frac{w^3}{3} 
-4\frac{\lambda^3}{\psi''\left( \frac{\gamma}{2}\right)}  \sum_{p\geq 2}\lambda^{-(2p+1)}\left(  \sum_{q=0}^{p} \xi_{p,q} k^{2q} w^{2p+1 - 2q} \right)\right)\\
&=  \int_{-\infty}^\infty dq\, \Ai(q)\int_0^\infty dv   \int_{C} \frac{dw}{ 2i\pi} \exp\left(    v( w -m)  +( q -  4 k^2) w
-\frac{4}{\psi''\left( \frac{\gamma}{2}\right)}  \sum_{p\geq 2} \lambda^{2-2p}\left(  \sum_{q=0}^{p} \xi_{p,q} k^{2q} w^{2p+1 - 2q} \right)\right)\\
&=  \int_{-\infty}^\infty dq\, \Ai(q)    \int_{C} \frac{dw}{ 2i\pi( w-m )} \exp\left(  ( q -   4 k^2) w   
- \frac{4}{\psi''\left( \frac{\gamma}{2}\right)}  \sum_{p\geq 2} \lambda^{2-2p}\left(  \sum_{q=0}^{p} \xi_{p,q} k^{2q} w^{2p+1 - 2q} \right)\right)\\
& =  \int_{-\infty}^\infty dq\, \Ai(q)   \exp\left(  ( q -   4 k^2 ) m   
-\frac{4}{\psi''\left( \frac{\gamma}{2}\right)}  \sum_{p\geq 2}\lambda^{2-2p}\left(  \sum_{q=0}^{p} \xi_{p,q} k^{2q} m^{2p+1 - 2q} \right)\right)\\
& =  \int_{-\infty}^\infty dq  \,\Ai\left(q + 4  k^2 \right)   \exp\left(  q  m   + O(\lambda^{-2}) \right).\\
\end{split}
\end{equation}
 Going back to the large-time equivalent of the quantity $\kernel_{10}$, we insert these representations, integrate 
 over the variables  $k_2$ with the constraint, and shift the variables of the two Airy functions, and take the large-$\lambda$ limit of the Gamma-function 
 factors and the saddle-point limit of the integral: 
\begin{equation}\label{Q1}
\begin{split}
Q_1(v_1,v_2) \underset{\lambda\to\infty}\sim 2\pi& \int_{-\infty}^\infty \frac{dk}{k}\int_{-\infty}^\infty dq \Ai\left(q + 4 k^2 \right)\int_{-\infty}^\infty dr \Ai\left( r + 4 k^2 \right) \\
&\times \int_{C} \frac{dm'}{2i\pi \lambda\sin(\pi m'/\lambda )}\exp\left( (q+r -2s - v_1 - v_2) m' +2i k( v_1 - v_2) +O(\lambda^{-2})  \right)\\
&\times \frac{\Gamma( 1  + 2ik_1/\lambda )\Gamma( 1+2ik_2/\lambda)}{\Gamma( 1-2ik_1/\lambda +m'/\lambda )\Gamma(1 + 2ik_2/\lambda + m'/\lambda)}\\
  \underset{\lambda\to\infty}\sim 2\pi &\int_{-\infty}^\infty \frac{e^{2i k( v_1 - v_2)}dk}{k}\int_{-\infty}^\infty dq \Ai\left(q + s +v_1 + 4 k^2 \right)\int_{-\infty}^\infty dr \Ai\left( r + s + v_2 + 4 k^2\right)  \theta(q+r ).
\end{split}
\end{equation}
  This expression is independent of the parameter of the log-gamma model, and reproduces
 the large-time limit of the corresponding kernel in the Lieb--Liniger model, were the time-evolution  factor 
 is determined by the energies of the string states in rapidity space rather than by the kernel $\Xi$ (see Eqs 142 and 182 of Section 7 \cite{CalabreseLeDoussalLong}).

%

%
\subsection{Two parity-invariant strings with independent numbers of particles}

 Analytic continuation to complex $m_i$ and $m_j$ is impractical for the zero-momentum sector. However, we can make use of the 
 expressions in terms of Gamma functions for the overlaps (Eq. 81 in \cite{CalabreseLeDoussalLong}). However, writing the 
  fixed-$m$ contribution factor as a Laplace transform as we did in the previous sector:
\begin{equation}
 \tilde{\Xi}(m,0)^t = \int_0^\infty \Upsilon( t,0,u) e^{-m u } du
\end{equation}
where $\Upsilon$ is an inverse Laplace transform, with the free energy shifted as prescribed in the paired-string  contribution:
\begin{equation}
\Upsilon( t,0,u)  = \int_{C} \frac{dz}{ 2i\pi}  \tilde{\Xi}(z,0)^t  e^{zu}.
\end{equation}
 The change of variable defined by  $u = \lambda v$ yields
\begin{equation}
 \tilde{\Xi}(m,0)^t = \lambda \int_0^\infty \Upsilon( t, 0,\lambda v) e^{-m\lambda v  } d v =  \lambda \int_0^\infty dv   \int_{C} \frac{dz}{ 2i\pi}  \tilde{\Xi}(z,0)^t  e^{z\lambda v} e^{-m\lambda v },
\end{equation}
and the change of variable $w = \lambda z$ puts the expression in a form suitable to a series expansion of the factor $\tilde{\Xi}$ around zero:
\begin{equation}
 \tilde{\Xi}(m,0)^t =  \int_0^\infty dv   \int_{C} \frac{dw}{ 2i\pi}  \tilde{\Xi}\left(\frac{w}{\lambda},0\right)^t  e^{v(w-\lambda m)}.
\end{equation}
 Let us insert the expansion of Eq. \ref{Taylorkm} with $k=0$,
 which, using  the time-scaling of the parameter $\lambda$, makes the time-independent  cubic term in the exponent appear, giving rise 
 to a factor that can be represented using the Airy function, with corrections that scale as negative powers of $\lambda^2$, 
 and contain all the dependence on the parameter $\gamma$ of the discrete polymer model:\\
 \begin{equation}\label{Q2Expansion}
\begin{split}
\tilde{\Xi}(m,0)^t &=   \int_0^\infty dv   \int_{C} \frac{dw}{ 2i\pi}  
 \int_{-\infty}^\infty dq \,  \Ai( q )  \exp\left( qw +   t \sum_{l\geq 2} \xi_{2l+1}   \left(\frac{w}{\lambda}\right)^{2l+1}  \right)    e^{v(w -\lambda m)},\\
& =  \int_{-\infty}^\infty dq \,  \Ai( q ) \int_{C} \frac{dw}{ 2i\pi}  \frac{1}{\lambda m - w}  \exp\left( qw  + t \sum_{l\geq 2} \xi_{2l+1}   \left(\frac{w}{\lambda}\right)^{2l+1}  \right)\\
& =  \int_{-\infty}^\infty dq \,  \Ai( q ) \exp\left( \lambda m q  +O(\lambda^{-2}) \right)
\end{split}
\end{equation}
  For integers values of the expansion parameter in the replica approach, we therefore obtain the following  moment 
 formula for the contribution of zero-momentum strings (with $\tilde{\lambda}$ the value of $\lambda$ associated to time $(t+1)/2$):\\
\begin{equation}
\begin{split}
Q_2^{mom}(v_1,v_2) & =\frac{(2\pi)^2}{4}   \int_{-\infty}^\infty dq \,  \Ai( q + s + v_1 )  \int_{-\infty}^\infty dr \,  \Ai( r + s + v_2 ) \\ 
 &\sum_{ m_1\geq 1}\sum_{ m_2\geq 1}\frac{e^{-\tilde{\lambda} m_1 q}}{m_1!}\frac{e^{-\tilde{\lambda} m_2 r }}{m_2!}(-1)^{{\mathrm{min}}(m_1,m_2)}{\mathrm{sgn}}(m_1-m_2) e^{\tilde{\lambda^3}( R_\gamma(m_1 ) + R_\gamma(m_2 ) )}\\
 & = \frac{(2\pi)^2}{4}   \int_{-\infty}^\infty dq \,  \Ai( q + s + v_1 )  \int_{-\infty}^\infty dr \,  \Ai( r + s + v_2 )  F( 2e^{-\tilde{\lambda} q},  2e^{-\tilde{\lambda} r} ) \left( 1 + O( {\tilde{\lambda}}^{-2})\right)
\end{split}
\end{equation}
As the function $F$ has been obtained through Borel transform and inverse Laplace transform, and expressed 
 in terms of Bessel functions in Appendix F of \cite{CalabreseLeDoussalLong} in terms of Bessel functions, we can conjecture
 that the analytically continued  large-time limit (or the time-independent component of it)  is captured by neglecting the 
  corrections in the above formula in a saddle-point approximation (which is consistent with the 
 expected Fredholm-determinant structure and the fact that these corrections were neglected when working out 
 the  one-string contribution), and taking the large-time limit worked out in \cite{CalabreseLeDoussalLong}:\\
\begin{equation}
   F( 2e^{-\tilde{\lambda} q},  2e^{-\tilde{\lambda} r} ) \underset{\lambda\to\infty}\sim\theta( r+q)( \theta(r)\theta( -q) - \theta( -r) \theta( q) )
\end{equation}
 so that
\begin{equation}
Q_2(v_1,v_2) \underset{\lambda\to\infty}\sim \frac{(2\pi)^2}{4}   \int_{-\infty}^\infty dq \,  \Ai( q + s + v_1 )  \int_{-\infty}^\infty dr \,  \Ai( r + s + v_2 ) \theta( r+q)( \theta(r)\theta( -q) - \theta( -r) \theta( q) ),
\end{equation}
 which together with the large-time limit of $Q_1$ and the Fredholm Pfaffian 
 structure of the generating function, allows to reproduce the argument of Section 7 of \cite{CalabreseLeDoussalLong}
 yielding the $\gamma$-independent limit of the two-string contribution to the generating function
\begin{equation}
 \lim_{t\to\infty} Z(n_s = 2,u) = 
\int   \det[ B_u(x_1,x_2) ]  dx_1 dx_2,\;\;\;
 \end{equation}
 where $B_u$ is the kernel based on the Airy function announced in Eq. \ref{kernel} (whose trace 
 can  be recognised in $Z(n_s=1, u)$),
 and to confirm through algebraic reasoning that at each order $n_s$ in the string expansion,
 \begin{equation}
 Z(n_s,u) \underset{\lambda\to\infty}\sim (-1)^{n_s}\int_{\mathbf{R}^{n_s}}\left( \prod_{k=1}^{n_s} dx_k \right) \det [ \theta( x_i)\Ai( x_i+x_j+u) \theta(x_j)]_{n_s\times n_s},
\end{equation}
 which yields the large-time convergence of the distribution of the rescaled free energy to the GOE Tracy--Widom distribution.

\section{Numerical tests}

As the above derivation contains several conjectural steps, independent 
 numerical checks are  crucial  to  compare  direct simulations at small system size to the predicted
   features of the large-time limit of the  model. 
  The low values of the parameter $\gamma$
 are the ones for which the divergence problem of the moments is the most severe. We 
  may therefore  choose the value $\gamma = 3$ for numerical evaluations, as in \cite{logGammaTLD},
 where the first two cumulants where studied for that value in the fixed-end case,
 for $N=10^4$ samples, and $t=4,096$, with a period $L=1,000$ and the starting point of the polymer at $(x=0,t=0)$.\\

\begin{figure}
\begin{center}
\includegraphics[width=6.5in,keepaspectratio]{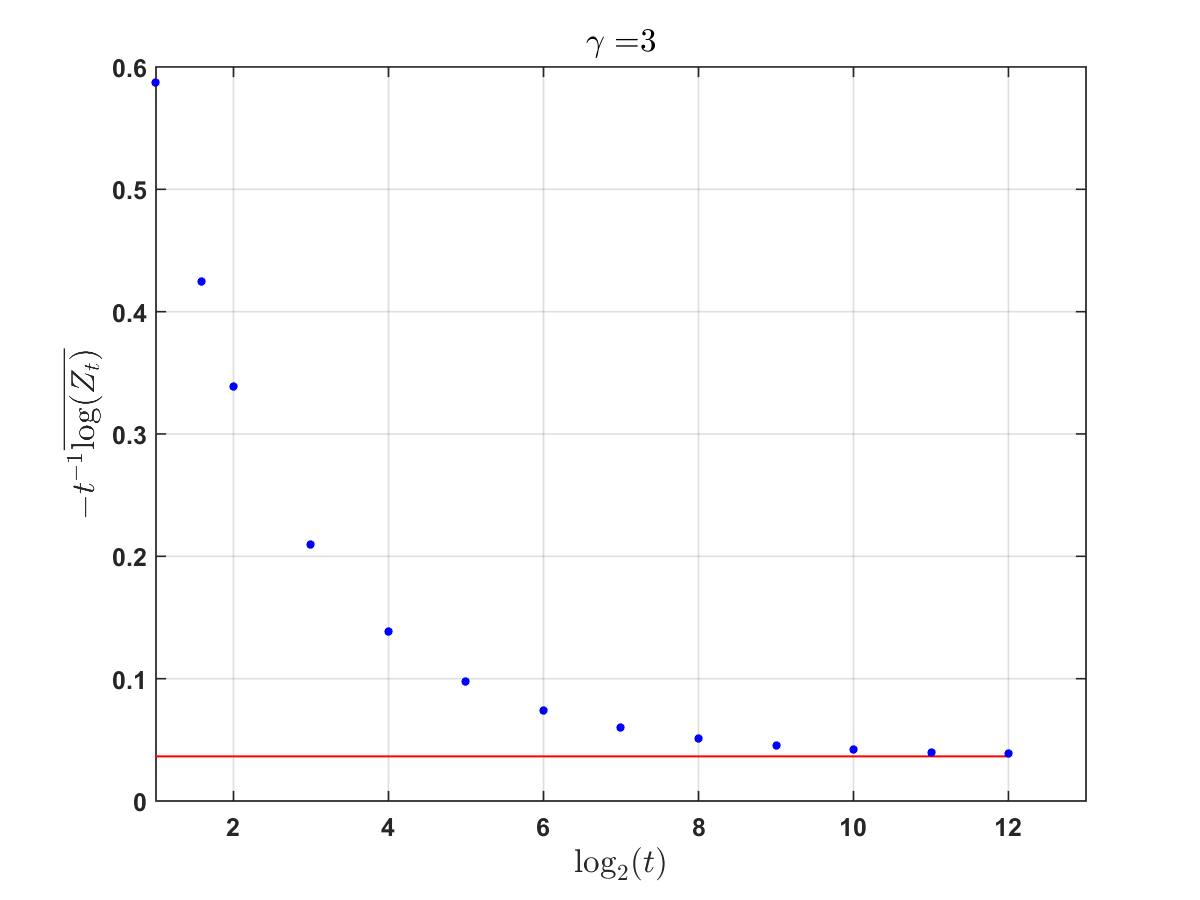}
\end{center}
\caption{ Convergence of the rescaled mean free energy  (blue dots, with mean taken over  $10^4$ samples)
 at $\gamma=3$ to the predicted  value $\psi( \gamma/2)\simeq 0.0365$ (red line).}
\label{figCumulFirst}
\end{figure}

 The non-universal ($\gamma$-dependent) predicted value 
\begin{equation}
 \lim_{t\to\infty} \left(  \frac{\overline{-\log Z_t}}{t} \right) = \psi\left( \frac{\gamma}{2}\right),
\label{cumulFirstLim}
\end{equation}
  appears as a horizontal asymptote on Fig. \ref{figCumulFirst}, where the l.h.s. of 
 Eq. \ref{cumulFirstLim} is plotted as a function of time at exponentially growing 
 values.\\

\begin{figure}
\begin{center}
\includegraphics[width=6.5in,keepaspectratio]{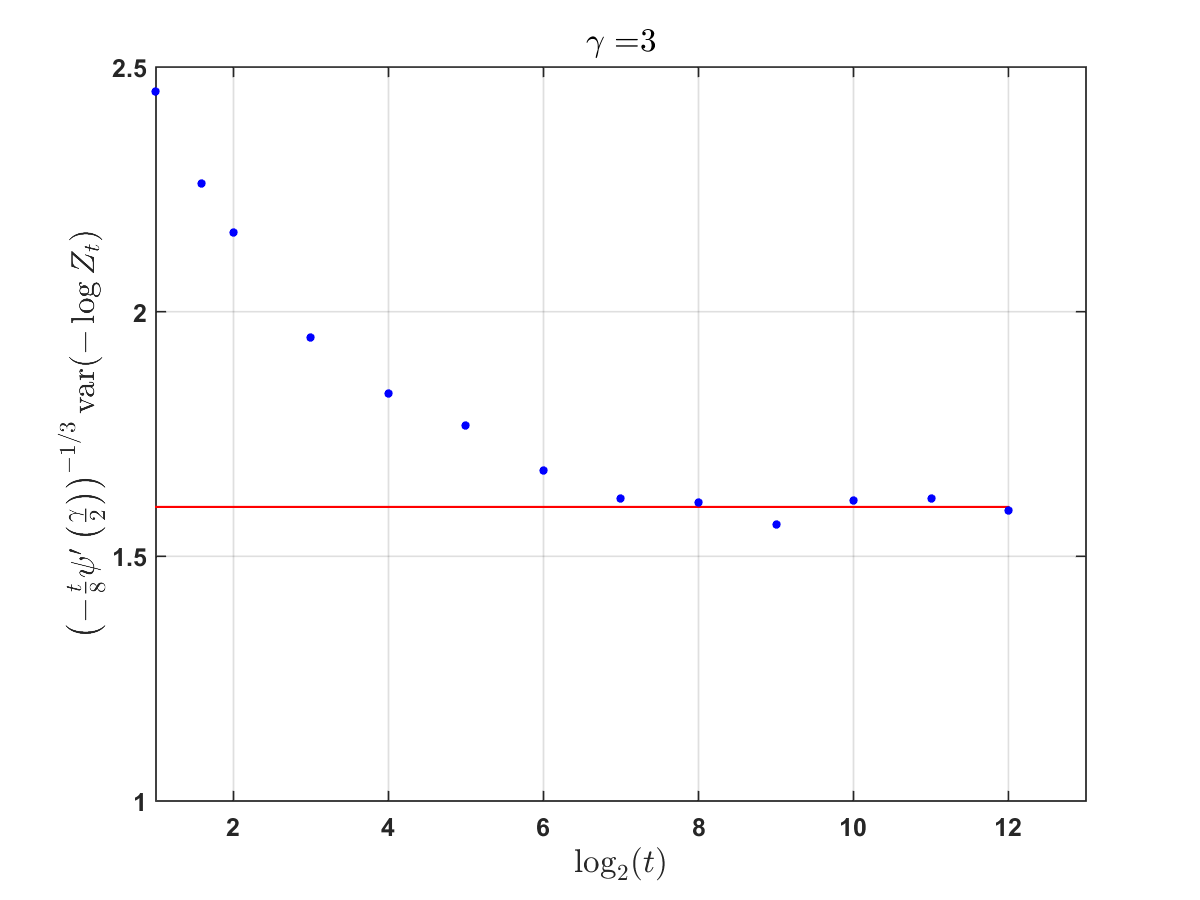}
\end{center}
\caption{Convergence of the variance of the rescaled  free energy  (blue dots, with mean taken over  $10^4$ samples)
 at $\gamma=3$ to the variance of the GOE probability distribution function (red line).}
\label{figCumulSecond}
\end{figure}

 The next cumulant, with a rescaling factor containing the contribution of the 
 parameter $\gamma$, allows to compare numerical results to the variance of the GOE
 Tracy--Widom distribution:
\begin{equation}
 \lim_{t\to\infty} \mathrm{Var}\left(  \frac{ \log Z_t}{\left( \frac{-t}{8}\psi''\left( \frac{\gamma}{2}\right) \right)^{1/3}}  \right) \simeq 1.60.
\label{cumulFirstLim}
\end{equation}
 The l.h.s. is plotted on Fig. \ref{figCumulSecond}, together with the predicted asymptote.\\

Moreover, the empirical distribution function of the rescaled free energy 
 can be compared to the GOE prediction. Let us denote by $Z_t,k$  the partition function at time $t$ of the  $k$-th sample.
  The  empirical cumulative distribution function (e.c.d.f.)  of the rescaled  free energy
 \begin{equation}
 {\mathcal{P}}_{N,t}(s) = \frac{1}{N}\sum_{k=1}^N \mathbf{1}\left(   \frac{\log Z_{t,k} + \psi\left( \frac{\gamma}{2}\right) t}{\left( -\frac{t}{8}\psi''\left(\frac{\gamma}{2} \right)\right)^{\frac{1}{3}}} < s\right),
\label{ecdf}
\end{equation}
at growing values 
 of time against tabulated values of $F_1$ 
(using MATLAB files\footnote{in code available from\\ 
{\ttfamily{https://www.researchgate.net/publication/316454158{\textunderscore}Simulation{\textunderscore}of{\textunderscore}the{\textunderscore}log-gamma{\textunderscore}polymer}}} available from\\ {\hbox{{\ttfamily{http://www.wisdom.weizmann.ac.il/\textunderscore nadler/Wishart{\textunderscore}Ratio{\textunderscore}Trace/TW{\textunderscore}ratio.html}}} for numerical evaluations of $F_1$ based on \cite{Bornemann,Nadler}). 
   Curves quickly superpose and the empirical curve at time $t=4,096$ needs zooming to distinguish
 it from the universal prediction (see Fig. \ref{cdfOptical}). 
  The two curves can be disentangled numerically  by plotting the logarithm of the relative  
  discrepancy
 \begin{equation}
\frac{\delta \mathcal{P}_{N,t}(s)}{  \mathcal{P}_{N,t}(s)} = \frac{ \left|\mathcal{P}_{N,t}(s) - F_1(s) \right|}{\mathcal{P}_{N,t}(s) },
\label{discrepancyCDF}
\end{equation}
as a function of $s$ (see Fig. \ref{relativeCDF}). In the domain $s > -4.4$ (i.e. for values of $s$ corresponding 
 to more than $F_1(s)>0.0022$), the relative discrepancy is found to be less than one percent, and 
 the average value of the decimal logarithm of the discrepancy over the domain is  $-5.58$.\\


 \begin{figure}[!ht]
\centering
    \subfloat[\label{subfig-1:dummy}]{%
      \includegraphics[width=0.8\textwidth]{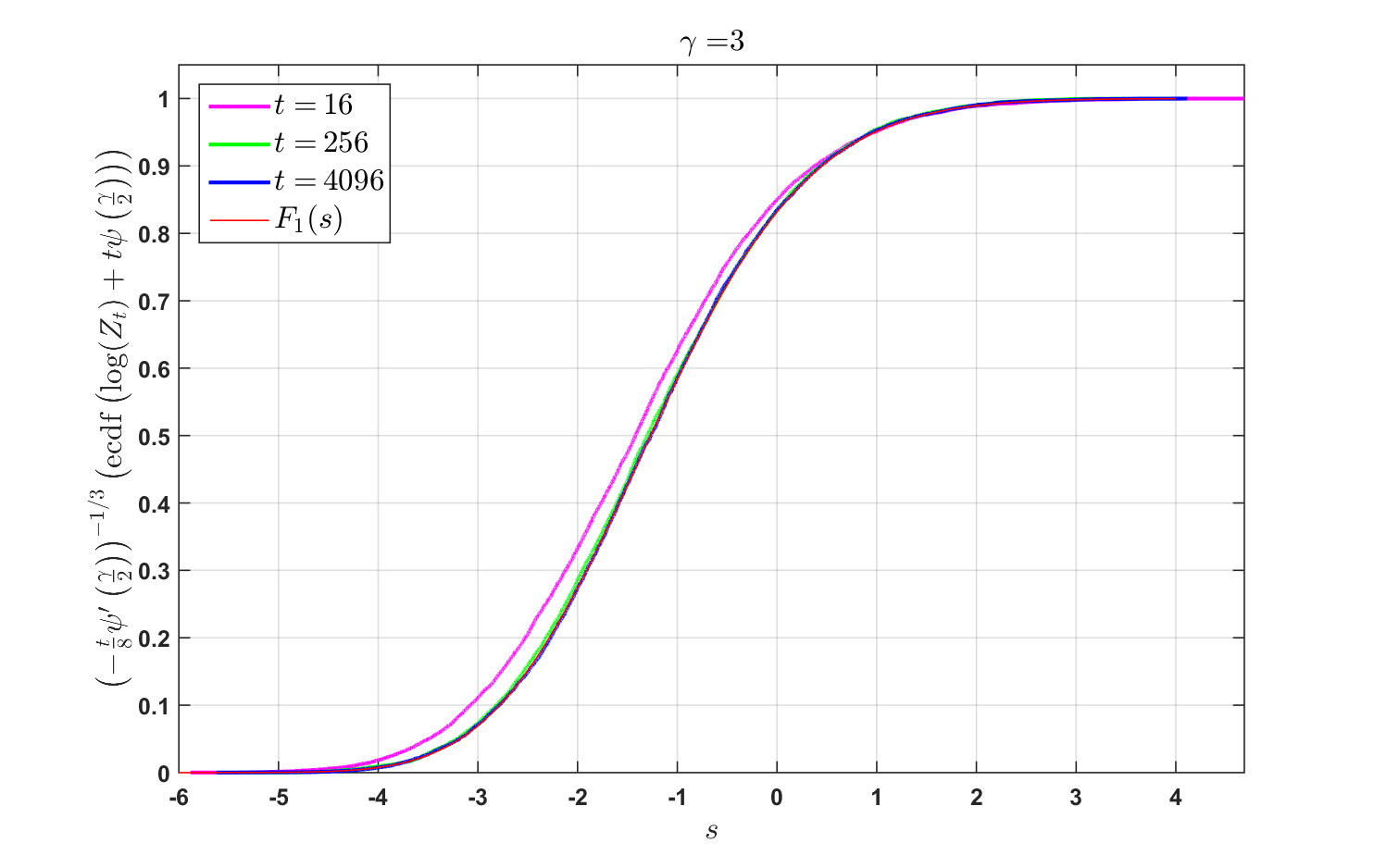}
    }
    \hfill
    \subfloat[\label{subfig-2:dummy}]{%
      \includegraphics[width=0.8\textwidth]{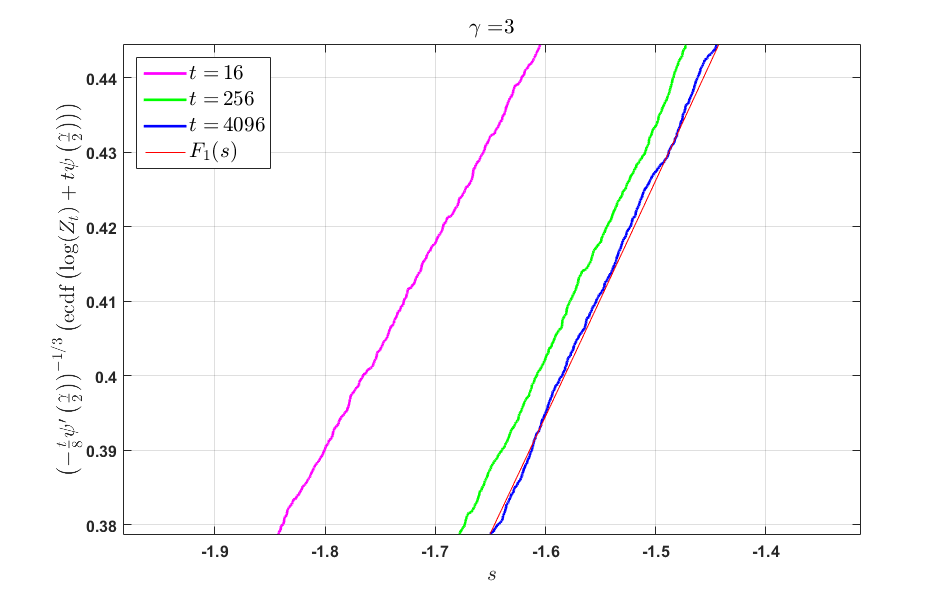}
    }
    \caption{(a) Empirical cumulative distribution function of the rescaled free energy (Eq. \ref{ecdf}) 
 for $t = 4,096$, based on $N = 10^4$ samples, blue curve), and GOE Tracy--Widom  function (red curve,
based on 4,097 tabulated values for regularly spaced values of $s$). (b) Zoom around the value of $s$ corresponding to the asymptotic value the second cumulant.}
    \label{cdfOptical}
  \end{figure}

\begin{figure}
\begin{center}
\includegraphics[width=7.5in,keepaspectratio]{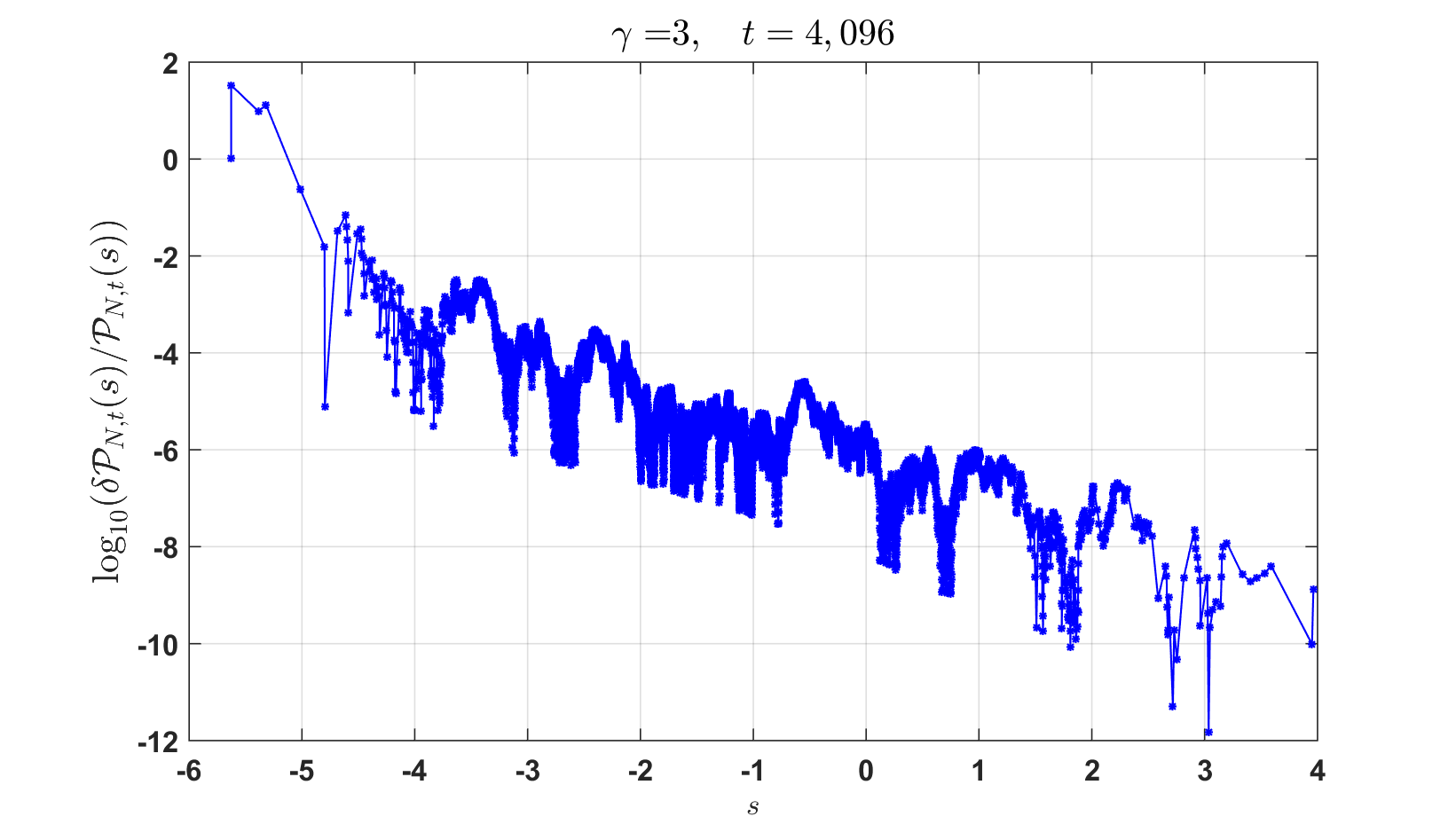}
\end{center}
\caption{ Logarithm of the relative discrepancy between the empirical cumulative distribution function of the rescaled free energy 
 and the universal GOE prediction (Eq. \ref{discrepancyCDF}),
   based on $N = 10^4$ samples, at $t = 4,096$.}
\label{relativeCDF}
\end{figure}

  Moreover, as the plotted numerical values of $F_1$ consist of $n_{F}=4097$ samples,
 the critical value of the absolute difference between the two curves in the
 two-sample Kolmogorov--Smirnov test  at level $\alpha = 0.001$ is 
 \begin{equation}
\delta = 1.95 \times \sqrt{\frac{N n_F}{N + n_F}} = 0.0362,
\end{equation} 
 whereas the maximum difference between the predicted curve and the empirical curve at $t=4,096$
  is found to be $0.0072$, which supports statistically the Tracy--Widom prediction.


%




%

\section*{Acknowledgments}
 This work 
 was supported by the Research Development Fund of Xi'an Jiaotong-Liverpool University (RDF-14-01-34).

\end{document}